\documentclass[journal=nalefd]{achemso}   

\usepackage[version=3]{mhchem} 
\usepackage{multirow}
\usepackage{soul}
\usepackage{array}
\usepackage[usenames,dvipsnames]{color} 
\usepackage[utf8]{inputenc}
\usepackage[T1]{fontenc}
\PassOptionsToPackage{hyphens}{url}
\usepackage{hyperref,url}
\usepackage{mathtools}
\usepackage{bm}
\usepackage{braket}
\usepackage{calrsfs}
\usepackage{multicol}
\usepackage{multirow}
\usepackage{makecell} 
\definecolor{mygreen}{rgb}{0.2,0.7,0.2}
\usepackage{dcolumn}
\usepackage{amsmath}
\usepackage{latexsym}
\usepackage{subcaption}
\usepackage{graphicx}
\usepackage{amssymb}
\usepackage{lineno} 
\usepackage{float}
\usepackage{physics}
\DeclareMathAlphabet{\pazocal}{OMS}{zplm}{m}{n}

\author{M. A. Garc\'ia-Bl\'azquez}
\email{manuelantonio.garcia@estudiante.uam.es}
\affiliation{~Departamento de F\'\i sica de la Materia Condensada, Universidad Aut\'onoma de Madrid, E-28049 Madrid, Spain}

\author{W. Dednam}
\affiliation{Department of Physics, Science Campus,
University of South Africa, Florida Park, Johannesburg 1710, South Africa}

\author{J. J. Palacios}
\affiliation{~Departamento de F\'\i sica de la Materia Condensada, Universidad Aut\'onoma de Madrid, E-28049 Madrid, Spain}
\alsoaffiliation{Condensed Matter Physics Center (IFIMAC), Universidad Autónoma de Madrid, E-28049 Madrid, Spain}

\title{Non-equilibrium spin accumulation and magneto-conductance in chiral nanojunctions from density-functional \& group theory}

\keywords{magneto-conductance, spin density, non-equilibrium quantum transport, chirality-induced spin selectivity, DFT calculations}

\begin{document}
\graphicspath{{Images/}}

\begin{tocentry}
\includegraphics[width=8.25cm,height=4.45cm]{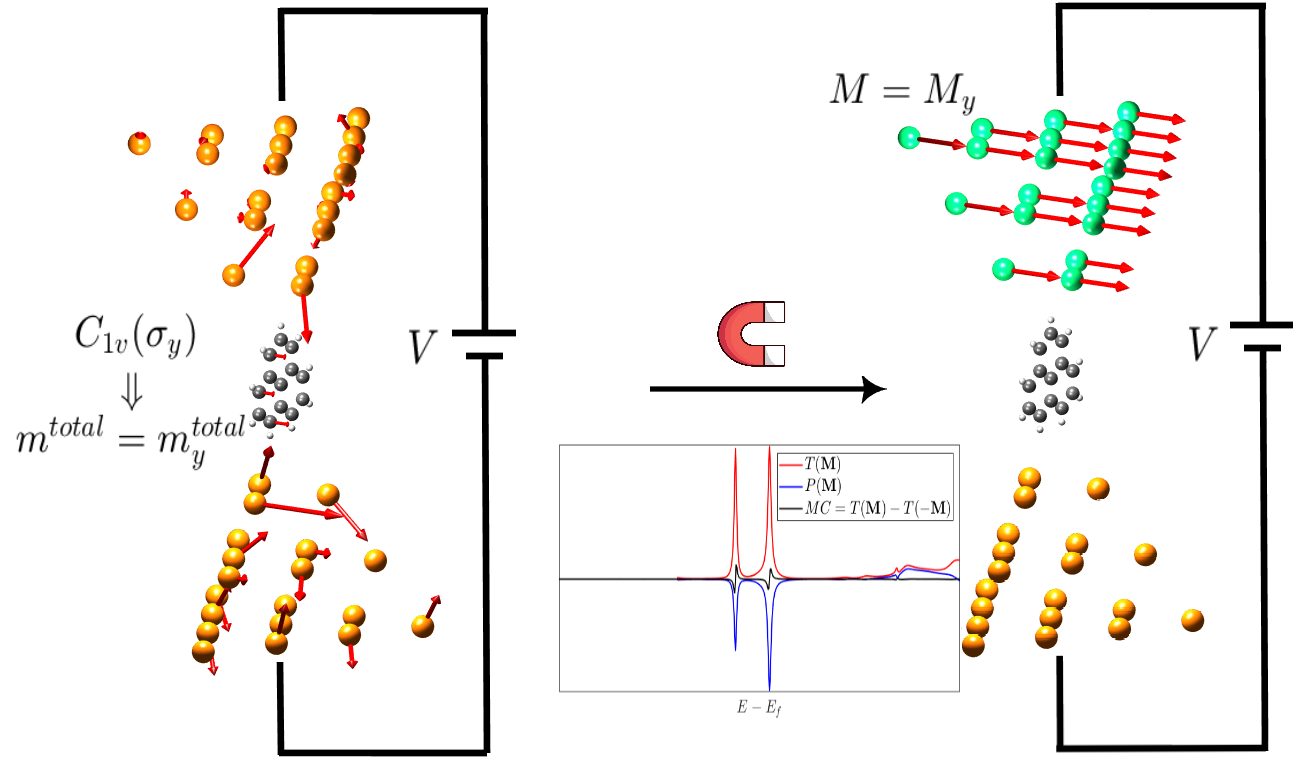}
\end{tocentry}


\begin{abstract} 
It is theoretically well established that a spin-dependent electron transmission generally appears in chiral systems, even without magnetic components, as long as a strong spin-orbit coupling is present in some of its elements. However, how this translates into the so-called chirality-induced spin selectivity in experiments, where the system is taken out of equilibrium,  is still debated. Aided by non-equilibrium DFT-based quantum transport calculations, here we show that, when spatial symmetries that forbid a finite spin polarization in equilibrium are broken, a \textit{net} spin accumulation appears at finite bias in an arbitrary two-terminal nanojunction. Furthermore, when a suitably magnetized detector is introduced in the system, the net spin accumulation, in turn, translates into a finite magneto-conductance. The symmetry prerequisites are mostly analogous to those for the spin polarization at any bias, with the vectorial nature given by the direction of magnetization. 
\end{abstract}

\flushbottom
\maketitle

\thispagestyle{empty}

\section{Introduction}
Relativistic effects experienced by electrons propagating through matter, most importantly in the presence of heavy atoms, are essential for many of the intrinsic magnetic properties of a variety of systems. In particular, in combination with the breaking of inversion symmetry in bulk materials, spin-orbit coupling (SOC) translates into the appearance of spin textures in reciprocal space and enables non-equilibrium phenomena such as spin-charge conversion, spin accumulation, and magneto-conductance (MC) \cite{1971JETPL..13..467D,silsbee2004spin,soumyanarayanan2016emergent,Calavalle2022}. In solids, the spin Hall effect \cite{Sinova.PhysRevLett.92.126603,Valenzuela2006} or the Edelstein effect \cite{EDELSTEIN1990233}, among others, result from it. More recently, chirality-induced spin selectivity (CISS) \cite{ray1999asymmetric,gohler2011spin,Naaman2012,Naaman2019,naaman2020chiral}, a seemingly related effect by which electrons propagating through chiral junctions (often involving chiral molecules) get spin polarized on average, has been proposed as an enabler for spintronics and quantum computing applications \cite{chiesa2023chirality,Aiello2022,naaman2020chiralB,forment2022chiral}. The CISS effect is typically revealed as a finite MC in two-terminal devices by introducing a ferromagnetic detector \cite{acsnano.0c07438,Maslyuk2018,fransson2022chiral}.
 
Over the years, a number of studies in the limit of zero bias have already established the crucial role of SOC in the emergence of a finite spin-polarization of the transmitted electrons in chiral junctions \cite{gersten2013induced,Maslyuk2018,Zollner2020a,Zollner2020b,alwan2021spinterface,Liu2021}. The importance of the metallic contacts in molecular junctions stems then from the introduction of a sufficiently strong SOC but, notably, also from the modification of the symmetry of the system as a whole, thus enabling for a finite spin polarization along certain otherwise forbidden directions for the molecule or material alone \cite{doi:10.1021/acsnano.2c11410,Guo2016}. In this regard, rotational symmetries along the transport direction play the same role as mirror symmetries (whose absence is alluded to in the chirality term), depending on the specific direction of spin polarization. 

On the other hand, in order to observe actual charge or spin currents, the system must be driven out of equilibrium. In our case a bias voltage between the two terminals must be applied \cite{acsnano.2c07088,yang2019spin,yang2020detecting}. Fundamentally, restrictions on the conductance in the unitary scattering scheme for coherent transport forbid the existence of a finite MC in two-terminal devices at equilibrium \cite{buttiker1988symmetry,zhai2005symmetry}. This so-called Onsager relation does not hold, however, under out of equilibrium conditions where one would expect to observe the CISS effect, among other interesting spin phenomena. Nonetheless, self-consistent non-equilibrium calculations in this context have barely been explored so far \cite{naskar2022}, in contrast with their equilibrium counterparts.

In this work, we perform a series of SOC-corrected density functional theory (DFT) non-equilibrium Green's function (NEGF) quantum transport calculations with the aim of illustrating the interrelations between spin polarization, spin accumulation and MC, as predicted by representation theory. First we perform the latter analysis within the NEGF formalism, determining the possible directions of spin polarization, accumulation, and MC (for which the vector nature is given by the ferromagnetic direction) according to the symmetries of the complete junction (reflections, accounting for the notion of chirality, and rotations not permuting the electrodes, see Table \ref{table1}). Since the selection rules are essentially coincident, the spin-polarization, in or out of equilibrium, must be accompanied by the accumulation of a net spin density at finite bias, ephemeral though this may be \cite{acsnano.2c07088,naskar2022}; which can then interact with the magnetic elements yielding a finite MC. With these theoretical and computational results we aim at elucidating the non-equilibrium spin-dependent response of nanoscopic systems in connection with equilibrium studies. Our quantitative results may nevertheless need to be complemented by electron correlations \cite{acs.jpclett.9b02929,acs.nanolett.1c00183} or electron-phonon interactions, which are currently deemed essential to quantitatively explain the large measured values and their scaling with temperature \cite{PhysRevB.102.035431,PhysRevB.102.235416,PhysRevB.102.214303,dubi2022spinterface}.

\section{Out-of-equilibrium spin density}
\subsection{Basics of out-of-equilibrium charge density}
Consider a junction with open boundary conditions described by a spin-dependent Hamiltonian operator $\hat{H}(\bm{r})$ and its corresponding retarded Green's function
\begin{equation}\label{Green}
\lim_{\eta\to0^{+}}[(E+i\eta)\hat{I}-\hat{H}(\bm{r})]\hat{G}^{+}(E,\bm{r})=\hat{I}\:,
\end{equation}
where $\hat{I}$ is the identity operator. Extensive quantities for the system may be computed from the one-particle density operator $\hat{\rho}$ as $\langle\hat{A}\rangle=\Tr[\hat{\rho}\hat{A}]$.
Consider a finite, atom-centered set of basis functions $\varphi_{\mu}(\bm{r})=\varphi_{a,\lambda}(\bm{r})=\varphi_{\lambda}(\bm{r}-\bm{t}_{a})\equiv\bra{\bm{r}}\ket{a,\lambda}=\bra{\bm{r}}\ket{\mu}$ with overlap matrix $\mathbb{S}$, where $\bm{t}_{a}$ is the coordinate vector of atom $a$ and $\mu=(a,\lambda)$ is a general multi-index for the usual basis parameters ($a\rightarrow\:$atoms, $\lambda=\lambda(a)\rightarrow\:$shells and functions therein). The basis is doubled to account for the spin degree of freedom $s\in\set{\uparrow,\downarrow}$: $\varphi^{s}_{\mu}(\bm{r})=\varphi_{\mu}(\bm{r})\ket{s}\equiv\bra{\bm{r}}\ket{s,\mu}$, yielding the block diagonal overlap matrix $S=I_{2}\otimes\mathbb{S}$. Upon restriction to a spatial subsystem $D$, in particular the device or scattering region of the junction, a non-Hermitian projection \cite{soriano2014theory} is taken for consistency with the Mulliken population analysis \cite{szabo2012modern} (equivalent to setting $\hat{A}=\hat{I}$):
\begin{equation} \label{TrD}
\Tr_{D}[\hat{\rho}\hat{A}]=\Tr[\hat{P}_{D}\hat{\rho}\hat{A}\hat{P}_{D}]=\sum_{\mu\in D}\sum_{s,\mu'}(\rho S A)_{s,\mu;s,\mu'}\mathbb{S}_{\mu',\mu}.
\end{equation} 
Here the projector has been defined as $\hat{P}_{D}=\sum_{\mu\in D}\sum_{s,\mu'}\ket{s,\mu}(\mathbb{S}^{-1})_{\mu,\mu'}\bra{s,\mu'}$ \footnote{Throughout the text, summations of basis indices in an unspecified range (such as $\sum_{\mu}$) are meant to span the whole basis.} and $\rho,A$ are the matrix representations of $\hat{\rho},\hat{A}$ in the dual basis \cite{soriano2014theory}.

Setting $\hat{\bm{A}}=\hat{\bm{\sigma}}\otimes\hat{I}$ in \eqref{TrD}, with $\bm{\sigma}=(\sigma^{x},\sigma^{y},\sigma^{z})$ the Pauli vector, one then obtains the magnetization or spin density \footnote{For simplicity, we employ the notation $\hat{I}$ for the identity operator both including and excluding the spin space. $\hat{I}_{s}$ denotes the identity on spin space alone. We have also omitted the global factor $\hbar/2$ for brevity.}
\begin{equation} \label{density}
\bm{m}=
\sum_{\mu\in D}\sum_{s}(\rho\cdot(\bm{\sigma}\otimes\mathbb{S}))_{s,\mu;s,\mu} =2\sum_{\mu,\mu'}\bm{\rho}_{\mu,\mu'}\mathbb{S}_{\mu',\mu}
\end{equation} 
where in the last step we have expanded $\hat{\rho}(\bm{r})=\hat{I}_{s}\otimes\hat{\rho}^{0}(\bm{r})+\sum_{i}\hat{\sigma}^{i}\otimes\hat{\rho}^{i}(\bm{r})$ in the two-dimensional spin space, $\hat{I}_{s}$ being the identity there, and $\hat{\bm{\rho}}=(\hat{\rho}^{x},\hat{\rho}^{y},\hat{\rho}^{z})$. This corresponds to the total spin density in the device, which is the sum over the contribution of all atoms in that region. Each individual atomic spin density is thus evaluated as
\begin{equation} \label{atomicdensity}
\bm{m}_{a}=\sum_{\lambda,s}(\rho\cdot(\bm{\sigma}\otimes\mathbb{S}))_{s,a,\lambda;s,a,\lambda}
=2\sum_{\lambda,\lambda',a'}\bm{\rho}_{a,\lambda;a'\lambda'}\mathbb{S}_{a',\lambda';a,\lambda}
\end{equation}
At equilibrium (linear regime), the leads present the same electrochemical potential $\mu$ and the density operator $\hat{\rho}^{\text{eq}}$ can be related to the retarded and advanced [$\hat{G}^{-}=(\hat{G}^{+})^\dagger$] Green's functions as
\begin{equation}\label{rhoeq}
\hat{\rho}^{\text{eq}}(\bm{r})=\frac{i}{2\pi}\int[\hat{G}^{+}(E,\bm{r})-\hat{G}^-(E,\bm{r})]f(E-\mu)dE
\end{equation} where $f$ is the Fermi distribution function. As usual, the retarded Green's function matrix block (dual basis) in the device can be expressed in terms of the device Hamiltonian and  the lead self-energies $\Sigma_{A},\Sigma_{B}$ as 
\begin{equation}\label{G+}
G^{+}_{D}(E)=\lim_{\eta\to0^{+}}[(E+i\eta)S_{D}-H_{D}-\Sigma_{A}(E)-\Sigma_{B}(E)]^{-1}
\end{equation} 
where 
\begin{equation}\label{self-energy}
\Sigma_{C}(z)=[zS_{DC}-H_{DC}][zS_{C}-H_{C}]^{-1}[zS_{CD}-H_{CD}],
\end{equation}
for $C=A,B$. 

Upon application of a bias $V$ to the junction, a difference of electrochemical potentials $\mu_{B}-\mu_{A}=eV$ is established between the electrodes $A$ and $B$. The Fermi function in \eqref{rhoeq} is then updated to $f(E-\min(\mu_{A},\mu_{B}))$, and the corresponding density matrix for the out of equilibrium (nonlinear) regime is $\hat{\rho}(\bm{r})=\hat{\rho}^{\text{eq}}(\bm{r})+\hat{\rho}^{\text{neq}}(\bm{r})$, with \cite{taylor2001ab,Jacob2011}: 
\begin{equation} \label{rhoneq}
\hat{\rho}^{\text{neq}}(\bm{r})=\frac{1}{2\pi i}\int_{\min(\mu_{A},\mu_{B})}^{\max(\mu_{A},\mu_{B})}\hat{G}^{<}(E,\mu_{A},\mu_{B},\bm{r})dE 
\end{equation}
where $\hat{G}^{<}$ is the lesser Green's function, whose matrix representation (dual basis) in the device is 
\begin{equation}\begin{aligned}\label{G<}
&G^{<}_{D}(E,\mu_{A},\mu_{B})=\\
&iG^{+}_{D}(E)[f(E-\mu_{A})\Gamma_{A}(E)+f(E-\mu_{B})\Gamma_{B}(E)]G^{-}_{D}(E)
\end{aligned}\end{equation}
with $\Gamma_{C}=i(\Sigma_{C}-\Sigma_{C}^{\dagger})$. 

Note that the expression \eqref{density} for the spin density is similar to that of the spin polarization \cite{nikolic2005decoherence}, the latter employing the scattering density matrix describing the set of outgoing channels instead of the eigenfunctions in the device region.

\subsection{Symmetry rules for spin density}
The system is in general described by a spin-dependent Hamiltonian of the form
$\hat{H}(\bm{r})=\hat{I}_{s}\otimes\hat{h}^{0}(\bm{r})+\hat{\sigma}^{x}\otimes\hat{h}^{x}(\bm{r})+\hat{\sigma}^{y}\otimes\hat{h}^{y}(\bm{r})+\hat{\sigma}^{z}\otimes\hat{h}^{z}(\bm{r})$. Under any (active) transformation $g$ of the point group $\pazocal{G}$ of the whole system, the Hamiltonain remains invariant:
\begin{equation*}\begin{aligned}
&\hat{g}\hat{H}(\bm{r})=(\hat{g}\hat{I}_{s})\otimes\hat{h}^{0}(g^{-1}\bm{r})+\sum_{i}(\hat{g}\hat{\sigma}^{i})\otimes\hat{h}^{i}(g^{-1}\bm{r})=\\
&\hat{H}(\bm{r})=\hat{I}_{s}\otimes\hat{h}^{0}(\bm{r})+\sum_{i}\hat{\sigma}^{i}\otimes\hat{h}^{i}(\bm{r})
\end{aligned}\end{equation*} 
If follows immediately that $\hat{h}^{0}(g^{-1}\bm{r})=\hat{h}^{0}(\bm{r})$. Furthermore, noting that the components of the Pauli vector transform as components of an angular momentum, i.e. $\hat{g}\hat{\sigma}^{i}=\sum_{j}\pazocal{D}^{1,+}_{j,i}(g)\hat{\sigma}^{j}$ with $\pazocal{D}^{1,+}$ the weight-1 representation of $O(3)$ (here restricted to $\pazocal{G}$), which is even under inversion, it follows that 
$\hat{h}^{i}(g^{-1}\bm{r})=\sum_{j}\pazocal{D}^{1,+}_{j,i}(g)\hat{h}^{j}(\bm{r})$ since $\pazocal{D}^{1,+}(g)$ has real entries. In particular this is the case for the spin-orbit interaction, $h^{i}_{\text{SOC}}(\bm{r})\propto(\nabla V(\bm{r})\times\bm{p})_{i}$, but it must hold irrespective of which spin-dependent terms are present in the Hamiltonian (albeit these may reduce $\pazocal{G}$ to a proper subgroup, unlike SOC). Let $\hat{G}^{+}(\bm{r})=\hat{I}_{s}\otimes\hat{g}^{0}(\bm{r})+\sum_{i}\hat{\sigma}^{i}\otimes\hat{g}^{i}(\bm{r})$, where the $E$ variable has been omitted for brevity. Then, transforming by $g\in\pazocal{G}$ in \eqref{Green} and inspecting the scalar component ($\hat{I}_{s}$ term)
\begin{equation*}
[(E+i\eta)\hat{I}-\hat{h}^{0}(\bm{r})]\hat{g}^{0}(\bm{r})-\sum_{i}\hat{h}^{i}(\bm{r})\hat{g}^{i}(\bm{r})=\hat{I}
\end{equation*}
one concludes that $\hat{g}^{0}(g^{-1}\bm{r})=\hat{g}^{0}(\bm{r})$, $\hat{g}^{i}(g^{-1}\bm{r})=\sum_{j}\pazocal{D}^{1,+}_{j,i}(g)\hat{g}^{j}(\bm{r})$ also. 

In equilibrium, taking the spatial $i-$th component in \eqref{atomicdensity} and exploiting the invariance of the spatial integrals under orthogonal transformations $g\in\pazocal{G}$, one obtains the conditions followed by the spin texture:
\begin{equation}\label{densitysymmetry} \begin{aligned} 
&m_{a}^{i,\text{eq}}=2\sum_{\lambda,\lambda',a'}\rho^{i}_{a,\lambda;a'\lambda'}\mathbb{S}_{a',\lambda';a,\lambda}=\\
&2\sum_{j}\pazocal{D}^{1,+}_{j,i}(g)\sum_{\lambda,\lambda',a'}\rho^{j}_{ga,\lambda;ga',\lambda'}\mathbb{S}_{ga',\lambda';ga,\lambda}=\sum_{j}\pazocal{D}^{1,+}_{j,i}(g)m^{j,\text{eq}}_{ga}
\end{aligned}\end{equation}
where $ga$ labels the device atom located at $g\bm{t}_{a}$ (which must exist for all $g\in\pazocal{G}$), and we have used the unitarity of the representations of the basis functions $\varphi_{\mu}$ (which make them cancel in the $\lambda,\lambda'$ summations, irrespective of the specific orbital character) and the fact that $\hat{\rho}^{\text{eq,i}}$ transforms according to $\pazocal{D}^{1,+}$. We have also defined the device region as being invariant under $\pazocal{G}$, without loss of generality. Since $g\in\pazocal{G}$ yields a bijection between the set of device atoms and itself, taking the sum over $a\in D$ in \eqref{densitysymmetry} one obtains
\begin{equation} \label{densityrule}
m^{i,\text{eq}}=\sum_{j}\pazocal{D}^{1,+}_{j,i}(g)m^{j,\text{eq}}
\end{equation}
For the mirror planes $\sigma_{k}$; $\pazocal{D}^{1,+}_{j,i}(\sigma_{k})=\delta_{i,j}(2\delta_{i,k}-1)$ for any set of three orthogonal directions, so that $m^{i}=m^{i}\delta_{i,k}$. On the other hand, for a $2\pi/n-$rotation $C_{n,k}$ along the direction $k$, $\pazocal{D}^{1,+}(C_{n,k})$ coincides with the standard rotation matrix and, by applying \eqref{densityrule} with $C_{n,k}$ and $C_{n,k}^{-1}$; one concludes that $m^{i}=m^{i}\delta_{i,k}$. We refer to these spatial symmetries that do not permute the electrodes as longitudinal. The total spin density must therefore point, as a (pseudo-)vector quantity, perpendicular to any longitudinal plane of symmetry, and along the axis of any non-trivial rotation symmetry. In particular, the simultaneous presence of a plane and an axis of symmetry forces $\bm{m}$ to vanish identically. Notice that this result also holds for each individual atomic layer in the junction (or more precisely, for each subset of atoms that is invariant under the corresponding symmetry operation). Additionally, the spin density of each individual atom that is invariant under $g'\in\pazocal{G}$, i.e. $a=g'a$, is subject to the selection rule \eqref{densityrule} (for $g'$ only) of the total density.

The previous discussion in the equilibrium case, including \eqref{densitysymmetry} and \eqref{densityrule}, are analogous for the out of equilibrium case with one exception: the symmetry operations must be limited to those that do not permute the electrodes (longitudinal). The operations that permute the electrodes (transversal) are actually not symmetries anymore, due to the different normalization of the electronic density (in the DFT formalism) in both electrodes. The longitudinal symmetries form a subgroup of the original group without bias, namely $\pazocal{G}_{l}=\pazocal{G}\cap\pazocal{G}_{A}\cap\pazocal{G}_{B}$, where $\pazocal{G}_{C}$ is the point group of electrode $C$ (sharing its invariant point with $\pazocal{G}$). To prove the non-equilibrium case, we first note that the scalar and pseudo-vector transformation rules are preserved under multiplication, that is, $\hat{f}^{0}(g^{-1}\bm{r})=\hat{f}^{0}(\bm{r})$ and $\hat{f}^{i}(g^{-1}\bm{r})=\sum_{j}\pazocal{D}^{1,+}_{j,i}\hat{f}^{j}(\bm{r})$, with $(\hat{I}_{s}\otimes\hat{f}^{0}+\sum_{i}\hat{\sigma}^{i}\otimes\hat{f}^{i})\equiv(\hat{I}_{s}\otimes\hat{f}_{1}^{0}+\sum_{i}\hat{\sigma}^{i}\otimes\hat{f}_{1}^{i})(\hat{I}_{s}\otimes\hat{f}_{2}^{0}+\sum_{i}\hat{\sigma}^{i}\otimes\hat{f}_{2}^{i})$ and $\hat{f}^{0}_{1,2},\hat{f}^{i}_{1,2}$ transforming analogously. Then, by transforming in all matrix elements of \eqref{self-energy} by $g\in\pazocal{G}_{l}$, one obtains for the components in the $\hat{I}_{s},\hat{\sigma}^{i}$ decomposition of spin space; 
$\hat{\Sigma}_{C}^{0}(g^{-1}\bm{r})=\hat{\Sigma}_{C}^{0}(\bm{r})$, $\hat{\Sigma}_{C}^{i}(g^{-1}\bm{r})=\sum_{j}\pazocal{D}^{1,+}_{j,i}(g)\hat{\Sigma}_{C}^{j}(\bm{r})$, with $C=A,B$. Performing these operations in \eqref{G<} one then concludes that the spin components of $\hat{G}^{<}$ have the same transformation properties under longitudinal operations as those of $\hat{G}^{+}$. For transversal symmetries without bias, employing \eqref{rhoeq} and \eqref{rhoneq} and proceeding as above, one obtains a generalized transformation rule reversing the bias voltage. Combining it with that for longitudinal symmetries:
\begin{equation}\label{maneqL}
m^{i,\text{neq}}_{a}(V)=\sum_{j}\pazocal{D}^{1,+}_{j,i}(g)m^{j,\text{neq}}_{ga}(V),\;\;g\in\pazocal{G}_{l}
\end{equation}
\begin{equation}\label{maneqT}
m^{i,\text{neq}}_{a}(V)=\sum_{j}\pazocal{D}^{1,+}_{j,i}(g)m^{j,\text{neq}}_{ga}(-V),\;\;g\in\pazocal{G}-\pazocal{G}_{l}
\end{equation}

Therefore, the restrictions induced by longitudinal symmetries \eqref{maneqL} on the total spin density out of equilibrium are ultimately equal to those on the spin polarization, which can be derived with the unitary scattering formalism in the linear regime \cite{doi:10.1021/acsnano.2c11410}, and extended to the nonlinear regime noting that the Green's functions in the Caroli formula \eqref{Caroli} transform equally under longitudinal operations.  

On the other hand, under time-reversal symmetry $\hat{\Theta}\hat{H}(\bm{r})\hat{\Theta}^{-1}=\hat{H}(\bm{r})$, where $\hat{\Theta}=(\hat{\sigma}^{y}\otimes\hat{I})\hat{K}$ up to an arbitrary phase, and $\hat{K}$ acts as complex conjugation. Then, performing this anti-unitary transformation in the expression for a general Green's function of complex argument $z$,
\begin{equation*}
[z\hat{I}-\hat{H}(\bm{r})]\hat{G}(z,\bm{r})=\hat{I},
\end{equation*}
it follows that $\hat{\Theta}\hat{G}(z,\bm{r})\hat{\Theta}^{-1}=\hat{G}(z^{*},\bm{r})=\hat{G}^{\dagger}(z,\bm{r})$. In particular, $\hat{\Theta}\hat{G}^{+}(E,\bm{r})\hat{\Theta}^{-1}=\hat{G}^{-}(E,\bm{r})$, the advanced Green's function. Thus, for the equilibrium density matrix \eqref{rhoeq}, 
\begin{equation*}
\hat{\Theta}\hat{\rho}^{\text{eq}}\hat{\Theta}^{-1}=\hat{\rho}^{\text{eq}}=\hat{I}_{s}\otimes\hat{\Theta}\hat{\rho}^{0}\hat{\Theta}^{-1}+\sum_{i}(-\hat{\sigma}^{i})\otimes\hat{\Theta}\hat{\rho}^{i}\hat{\Theta}^{-1}
\end{equation*}
Then $\hat{\Theta}\hat{\rho}^{0}\hat{\Theta}^{-1}=\hat{\rho}^{0}$, $\hat{\Theta}\hat{\rho}^{i}\hat{\Theta}^{-1}=-\hat{\rho}^{i}$, and transforming in the scalar products of \eqref{atomicdensity}, one concludes that $\bm{m}_{a}=0$ for all atoms, and obviously $\bm{m}=0$. In the absence of magnetic elements (and fields) in the junction, the spin density at equilibrium is therefore locally vanishing. For the magnetization \eqref{atomicdensity} in the nonlinear regime, from \eqref{self-energy}, \eqref{G<} and the above, it follows that 
\begin{equation}\begin{aligned}
&\bm{m}_{a}=\frac{1}{2\pi}\sum_{\lambda,s}\left(\int G^{-}_{D}[f(E-\mu_{A})\Gamma_{A}+f(E-\mu_{B})\Gamma_{B}]G^{+}_{D}\cdot\right.
\\&\left.dE
(-\bm{\sigma}\otimes\mathbb{S})\right)_{s,a,\lambda;s,a,\lambda}
\end{aligned}\label{spindensity}\end{equation}
and an analogous result holds for the total spin density \eqref{density}. In contrast to the linear regime, Eq. \eqref{spindensity} generally does not result in an actual restriction on the spin density. 

Note that both time-reversal and the spatial operations are here formally defined as symmetries according to the single-particle Hamiltonian of the system, i.e., they are intrinsic to the system irrespective of the applied bias, which then modifies the effect of the operations that change the propagation ($\Theta$ and the transversal symmetries) in the NEGF framework. 

\subsection{Numerical results: spin density}
\begin{figure}[H]
\centering 
\begin{subfigure}{0.4\textwidth}
    \includegraphics[width=\textwidth,height=6cm]{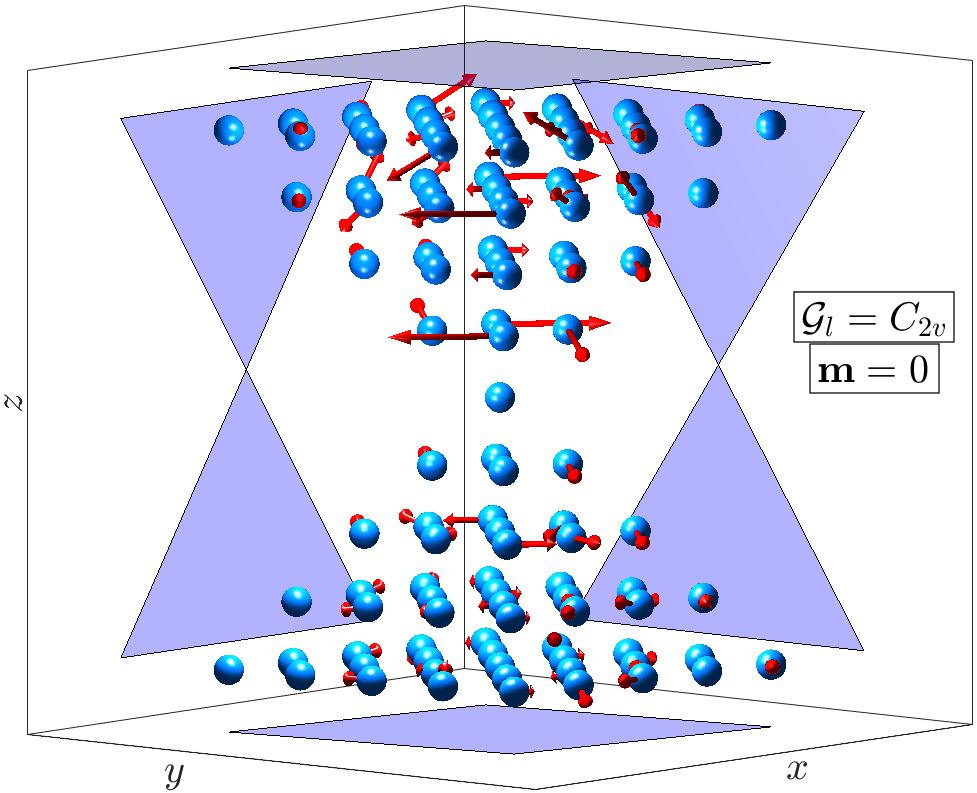}
    \caption{}
    \label{Fig1A}
\end{subfigure}
\begin{subfigure}{0.4\textwidth}
    \includegraphics[width=\textwidth,height=6cm]{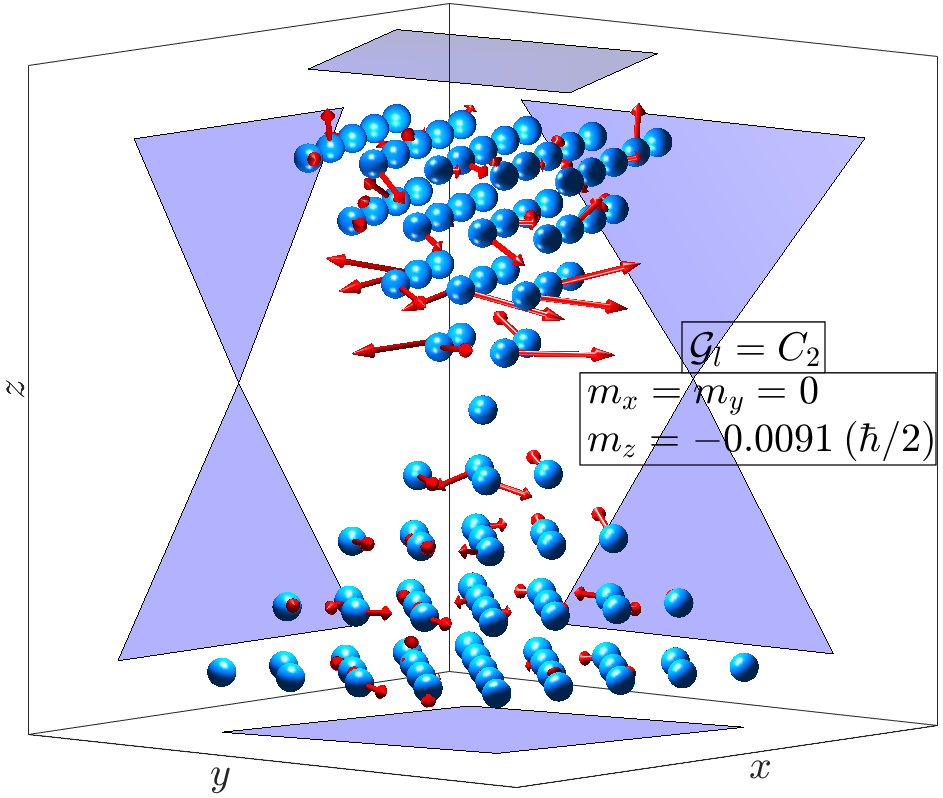}
    \caption{}
    \label{Fig1B}
\end{subfigure}
\begin{subfigure}{0.18\textwidth}
 \includegraphics[width=2.1\textwidth,height=4cm,angle=270]{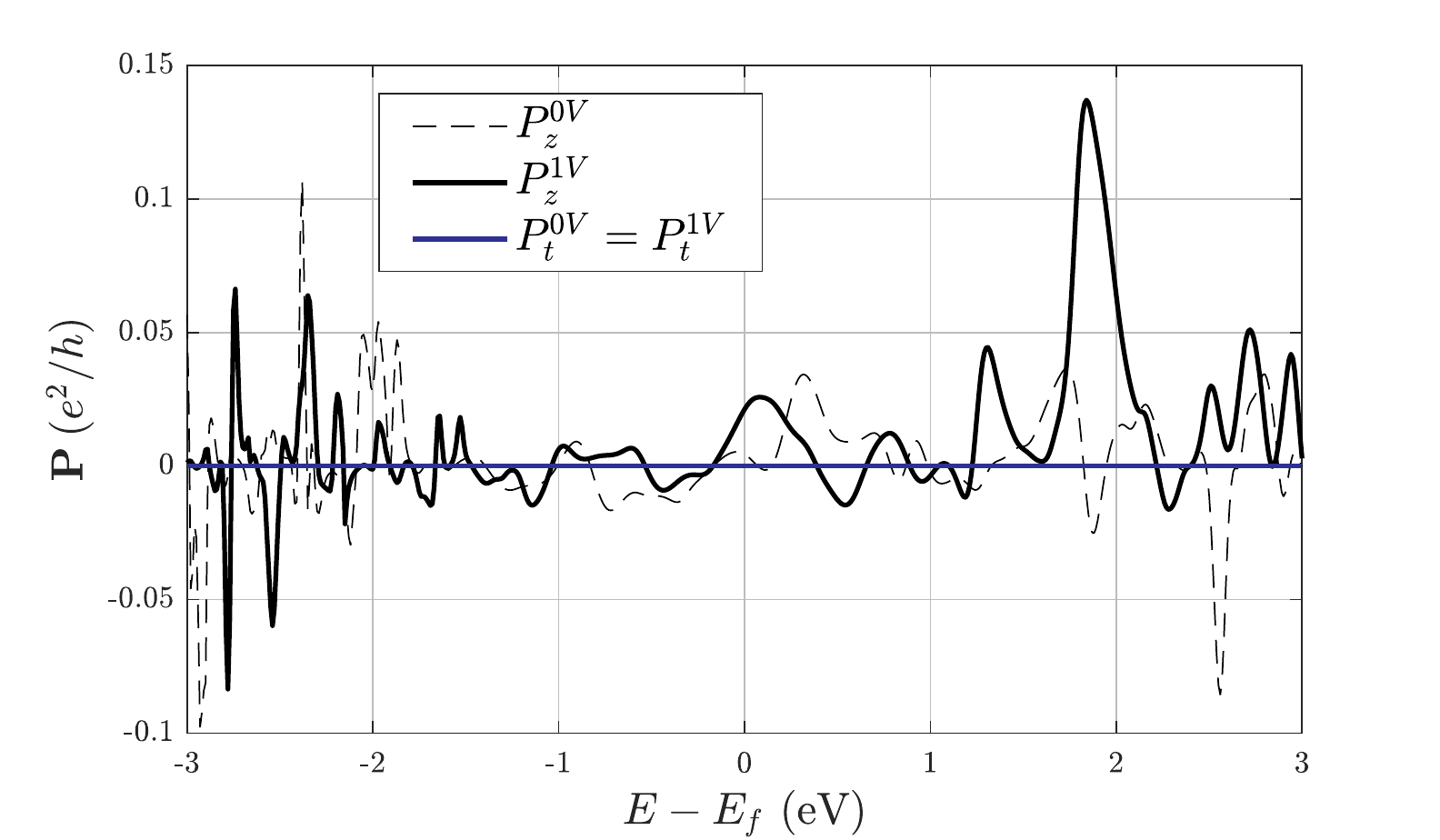}
    \caption{}
    \label{Fig1C}
\end{subfigure}
\begin{subfigure}{0.4\textwidth}
    \includegraphics[width=\textwidth]{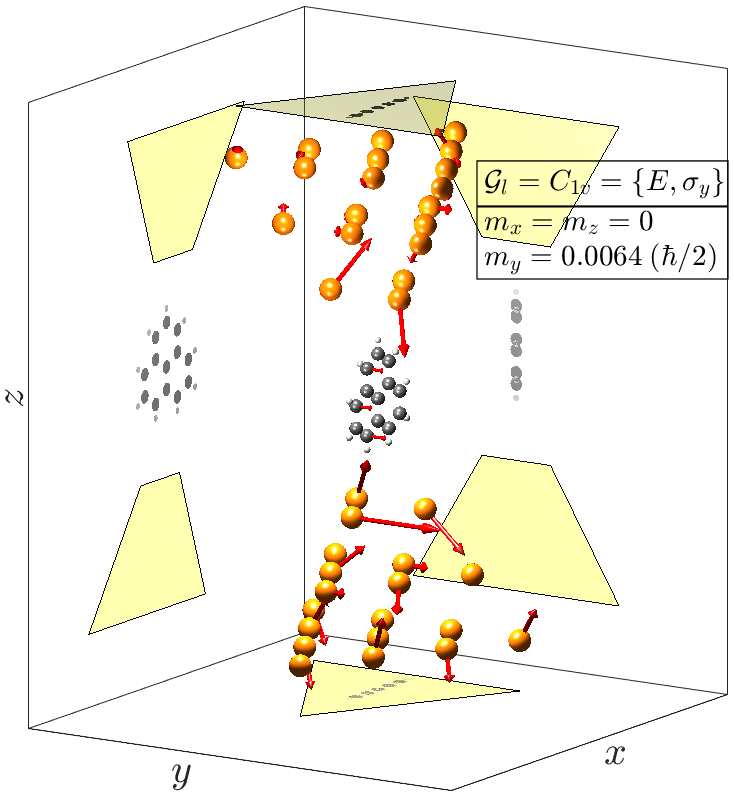}
    \caption{}
    \label{Fig1D}
\end{subfigure}
\begin{subfigure}{0.4\textwidth}
    \includegraphics[width=\textwidth]{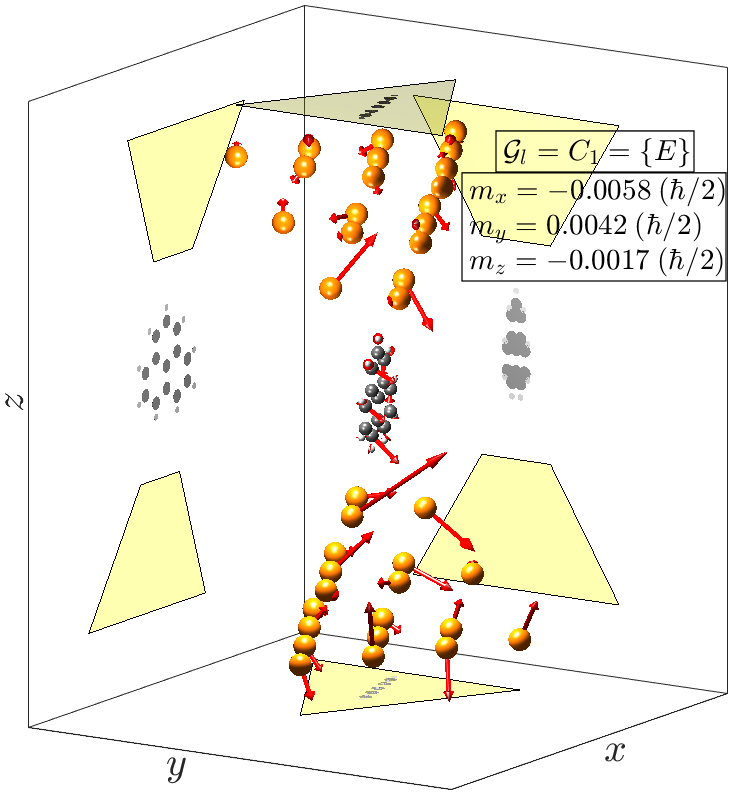}
    \caption{}
    \label{Fig1E}
\end{subfigure}
\begin{subfigure}{0.18\textwidth}    \includegraphics[width=2\textwidth,height=4cm,angle=270]{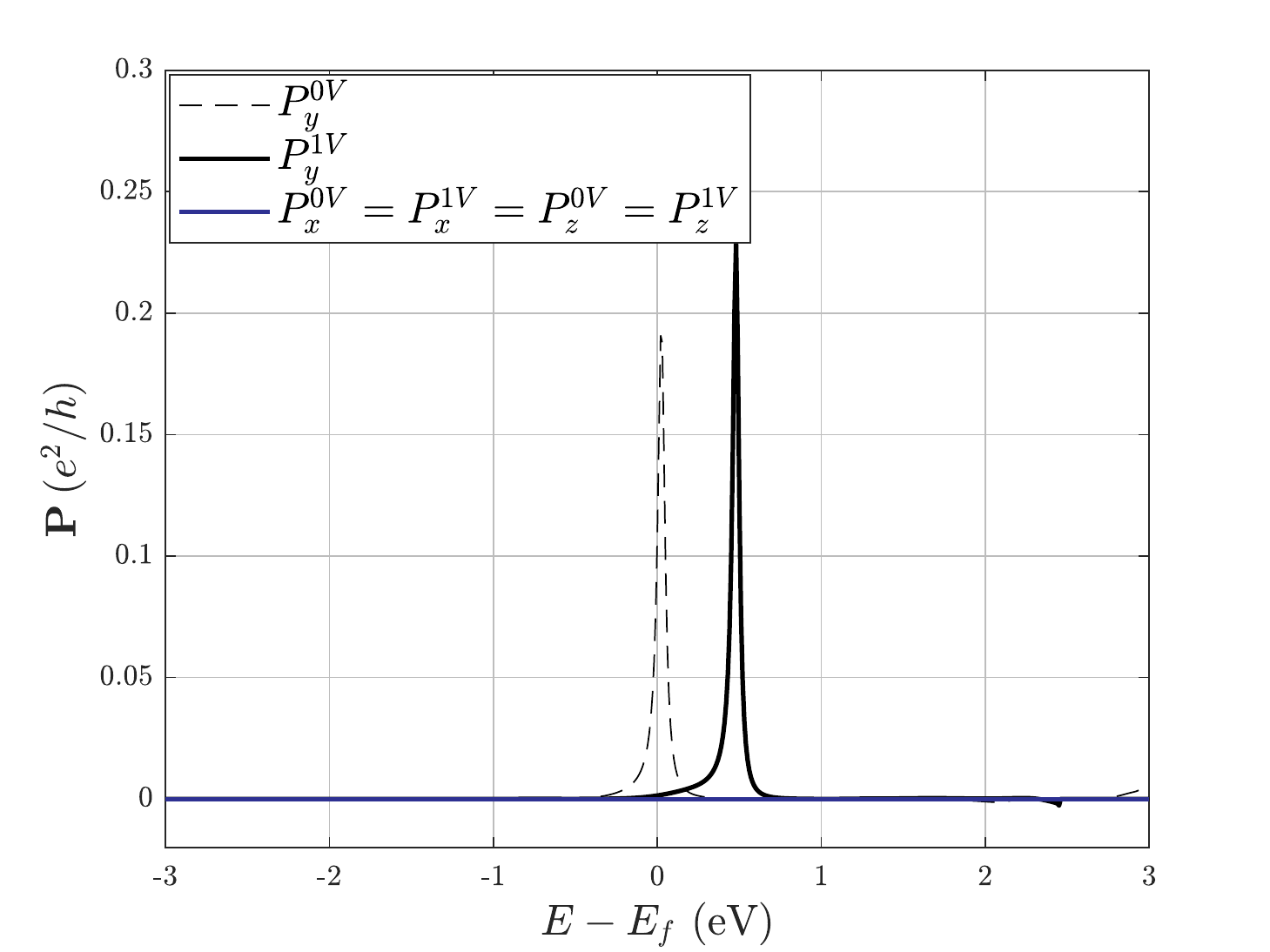}
    \caption{}
    \label{Fig1F}
\end{subfigure}
\caption{\label{Fig1} 
Non-equilibrium spin density per atom with bias $1V$. $x$, $y$ and $z$ projections are shown to help visualize the structures, with the top and bottom contacts having a separate projection above and below (resp.) the structure. The magnitude of the vectors is purely illustrative, and has been uniformly scaled by the same factor in (a,b) and (d,e) to make them clearly visible. (a) Bare W contacts aligned, point group $\pazocal{G}=D_{2h}$, group of longitudinal symmetries $\pazocal{G}_{l}=C_{2v}$. (b) Bare W contacts misaligned, $\pazocal{G}=\pazocal{G}_{l}=C_{2}$. (d) Pb contacts with aligned triangulene molecule, $\pazocal{G}=\pazocal{G}_{l}=C_{1v}$. (e) Pb contacts with misaligned triangulene molecule, $\pazocal{G}=\pazocal{G}_{l}=C_{1}$. (c,f) Equilibrium ($0V$) and non-equilibrium ($1V$) spin polarization components of, respectively, the systems in (b,d); where $P_{t}$ stands for any transversal direction ($\perp\bm{z}$). SOC is included only in the contacts, excluding the shown layers of maximum and minimum $z$.}
\end{figure}

We have performed a series of DFT-based non-equilibrium quantum transport calculations to compute the spin density and polarization in this regime (computational details can be found in the Methods section). Two illustrative cases are considered here: bare W nanocontacts, see Fig. \ref{Fig1A}-\ref{Fig1B}; and a molecular bridge composed of Pb electrodes and a triangulene molecule, see Fig. \ref{Fig1D}-\ref{Fig1E}. 

As can be observed, a finite spin density is accumulated per atom in the presence of a bias voltage (except if such an atom is invariant under both a longitudinal plane $\sigma_{l}$ and axis $C_{n,l}$, as are e.g. all atoms in a mono-atomic chain). In these non-magnetic systems, the spin density is a purely non-equilibrium quantity induced by the SOC. The net density is, however, strictly vanishing for aligned crystallographic electrodes (at least for those that are cubic, tetragonal or orthorhombic in the bulk), such as in Fig. \ref{Fig1A}, due to the presence of both symmetries $\sigma_{l},C_{n,l}\in\pazocal{G}_{l}$ as dictated by \eqref{maneqL}. Upon a relative rotation of a single contact, the mirror plane is generally removed and a net magnetization along the transport direction is hence enabled, see Fig. \ref{Fig1B}.

Likewise, in the molecular bridge of Fig. \ref{Fig1D} only $\sigma_{l}$ or, in the chosen coordinates system, $\sigma_{y}$ is a symmetry. Thus by \eqref{maneqL} the net spin density must point perpendicular to the plane: $\bm{m}=m_{y}\hat{\bm{y}}$. For each atom that lies in such a plane, in particular the whole molecule, the same rule applies individually. Note that the asymmetry of the electrodes is inherited in the spin accumulation of the molecule, which is by itself achiral with, in the present orientation with respect to the longitudinal direction; $C_{2v}=\pazocal{G}_{l}^{\text{molec}}\subset\pazocal{G}^{\text{molec}}=D_{3h}$ and would thus present a vanishing $\bm{m}$ if the geometry of the electrodes was ignored. Numerically, $\abs{\bm{m}_{a}}$ in the molecule can be comparable or larger than in other heavier atoms of the electrodes, even if the SOC strength of the former is negligible in comparison. In analogy with Fig. \ref{Fig1B}, the rotation of the molecule with respect to the contacts will generally break the longitudinal mirror symmetry, allowing for a finite net spin density in arbitrary directions, see Fig. \ref{Fig1E} (note that in this case there are no spatial symmetries at all). 
 
In Fig. \ref{Fig1C}-\ref{Fig1F} the polarization both in and out of equilibrium for systems \ref{Fig1B} and \ref{Fig1D} is depicted. As can be observed, the numerical behaviour of this quantity with the bias voltage is quite specific to the given system but, as expected, the symmetry rules enforcing the vanishing of any $\bm{P}$ components do remain valid in the nonlinear regime. Note that the finite components of net spin density and polarization, the latter either in or out of equilibrium, are indeed the same.

\section{Magneto-conductance}
\subsection{Symmetry rules}
The total transmission (or dimensionless conductance) in a system with a ferromagnetic component of macroscopic magnetization $\bm{M}$, namely $T_{\bm{M}}=T^{\uparrow\uparrow}_{\bm{M}}+T^{\uparrow\downarrow}_{\bm{M}}+T^{\downarrow\uparrow}_{\bm{M}}+T^{\downarrow\downarrow}_{\bm{M}}$,
can be computed from the spin-resolved Caroli formula at a given energy:
\begin{equation}\label{Caroli}
T^{s,s'}(E)=\sum_{\mu\in D}\left(\Gamma_{A}(E)G^{-}_{D}(E)\Gamma_{B}(E)G^{+}_{D}(E)\right)_{s,\mu;s',\mu}
\end{equation}
This expression is valid for both in and out of equilibrium situations, with differences in the Green's functions induced from the self-consistent Hamiltonian through the charge density. The difference of conductance for opposite magnetizations, $\Delta T_{\bm{M}}=T({\bm{M}})-T({-\bm{M}})$, is the so called magneto-conductance (MC). This quantity, when measured in experiments with chiral and ferromagnetic components, is often considered a manifestation of the CISS effect \cite{acsnano.0c07438,Maslyuk2018,fransson2022chiral}. Note that this differs from the tunnel magneto-conductance, for which the two magnetic configurations are not reversed as a whole. At equilibrium, a finite MC in two-terminal devices is forbidden by Onsager's reciprocity, which can be proved within the scattering formalism for coherent transport by performing the time-reversal operation (not a symmetry for finite $\bm{M}$) and invoking the unitarity of the scattering matrix \cite{zhai2005symmetry,doi:10.1021/acsnano.2c11410}. This formalism no longer holds in the nonlinear regime, allowing in principle for a finite MC if spatial symmetries do not forbid it. 

Consider a junction with a ferromagnetic component magnetized along the longitudinal direction, say $\bm{M}=M\bm{z}$, and which presents a longitudinal plane of symmetry $\sigma^{l}$ if magnetism is disregarded (or equivalently, $\Theta\sigma^{l}$ is an element of the magnetic point group, but not $\sigma^{l}$). Then the Hamiltonian operator of the system must transform as $\hat{H}_{\bm{M}}(\sigma^{l}\bm{r})=\hat{H}_{-\bm{M}}(\bm{r})$, so that applying this operation in the space integrals of \eqref{Caroli}, one obtains $T^{s,s'}_{\bm{M}}(E)=T^{\overline{s},\overline{s}'}_{-\bm{M}}(E)$ with $\overline{s}$ the opposite spin state of $s$ \cite{doi:10.1021/acsnano.2c11410}. Therefore, in this case $\Delta T_{\bm{M}}(E)=0$ even out of equilibrium. 
Similarly, when the magnetization $\bm{M}$ is along an arbitrary transversal direction, i.e., perpendicular to the transport direction, a longitudinal $2$-fold rotation symmetry $C_{2,z}$ (when disregarding magnetism) has the same effect as $\sigma_{l}$ above. That is, the MC for transversal ferromagnets must vanish if $\Theta C_{2,z}$ is a (anti-unitary) symmetry of the system, even in the nonlinear regime. Note that this argument does not hold for other rotations $C_{n,z}$, $n\geq3$ ($\Theta C_{n,z}$ is not an anti-unitary symmetry), since then the results for $C_{n,z}$ and $C_{n,z}^{-1}$ yield different $\bm{M}$ vectors and cannot be combined.

Importantly, these selection rules for the MC are analogous to those for both the total spin density and the spin polarization (see next paragraph) as discussed above, with the direction of the magnetization $\bm{M}$ playing the same role as the spatial component of $\bm{m}$ or $\bm{P}$. This is consistent with the expectation of detecting the net spin density along a given direction by introducing a ferromagnet oriented along that direction. As explained above, the only exception to this shared selection rules occurs for transversal directions in systems whose point group has a main (longitudinal) rotation axis of odd order, e.g. $\pazocal{G}=C_{3}$ forces the vanishing of $P_{t}$ and $m_{t}$ along any transversal direction but not of $\Delta T_{M_{t}}$ (see numerical example in the Supporting Information). 

It may be worth noting that the introduction of the ferromagnet induces a finite spin polarization $\bm{P}(\bm{M})$ regardless of whether the underlying $\bm{P}(0)$, which is induced by SOC if spatial symmetries allow for it, is vanishing or not. The quantity that actually shares the selection rules with $\Delta T_{\bm{M}}$, $\bm{m}$ and $\bm{P}(0)$ (except for odd-order rotations) is the symmetrized $\Delta\bm{P}_{\bm{M}}=\bm{P}(\bm{M})+\bm{P}(-\bm{M})$.

\begin{table}[h!]\caption{\label{table1} Selection rules. Subscript $l$ means longitudinal, i.e., not permutting the electrodes.}\centering\begin{tabular}{| c | c | c | c | c | c |} 
\hline
Symmetry & $\bm{m}_{a}^{\text{eq}}$ & $\bm{m}^{\text{eq}}$ & $\bm{m}_{a}^{\text{neq}}$ & $\bm{m}^{\text{neq}}$ & $\Delta T_{\bm{M}}$\\[0.08cm] \hline
$\Theta$ & 0 & 0 & & & Not a symmetry \\[0.08cm] \hline
$\sigma$ & \makecell{$\perp\sigma:\bm{m}_{a}^{\text{eq}}=\bm{m}_{\sigma a}^{\text{eq}}$\\$\parallel\sigma:\bm{m}_{a}^{\text{eq}}=-\bm{m}_{\sigma a}^{\text{eq}}$} & $\perp\sigma$ & \makecell{$\perp\sigma_{l}:\bm{m}_{a}^{\text{neq}}=\bm{m}_{\sigma_{l}a}^{\text{neq}}$\\$\parallel\sigma_{l}:\bm{m}_{a}^{\text{neq}}=-\bm{m}_{\sigma_{l}a}^{\text{neq}}$\\only for $\sigma_{l}$} & $\perp\sigma_{l}$ & \makecell{$\Delta T_{\bm{M}\perp\sigma_{l}}$\\[0.1cm]($\Theta\sigma_{l}$ symm.) } \\[0.08cm] \hline
$C_{n}$ & \eqref{densitysymmetry} & $\parallel C_{n}$ & \makecell{\eqref{maneqL}\\only for $\sigma_{l}$} & $\parallel C_{n,l}$ & \makecell{$\Delta T_{\bm{M}\parallel C_{n,l}}$, \\[0.1cm] only for $n=2$\\[0.1cm] ($\Theta C_{2,l}$ symm.)}  \\[0.08cm] \hline
\end{tabular}\end{table}

\subsection{Numerical results: magneto-conductance}
\begin{figure}[H]
\centering 
\begin{subfigure}{0.49\textwidth}
\includegraphics[width=\textwidth,height=6cm]{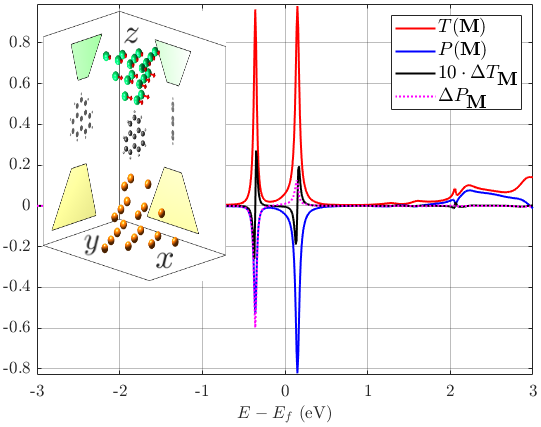}
    \caption{}
    \label{Fig2A}
\end{subfigure}
\begin{subfigure}{0.49\textwidth}
\includegraphics[width=\textwidth,height=6cm]{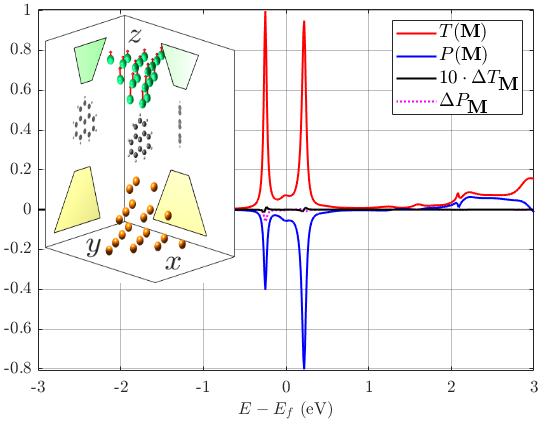}
    \caption{}
    \label{Fig2B}
\end{subfigure}
\begin{subfigure}{0.49\textwidth}
\includegraphics[width=\textwidth]{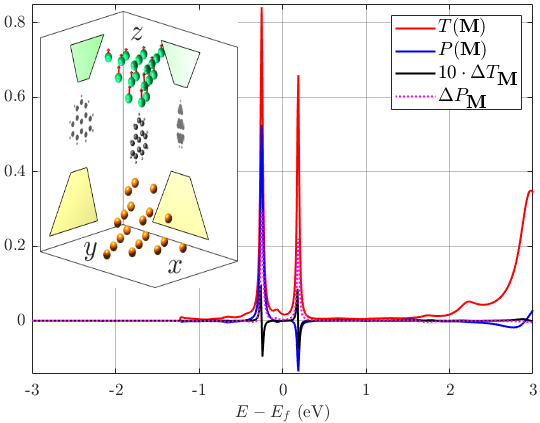}
    \caption{}
    \label{Fig2C}
\end{subfigure}
\begin{subfigure}{0.49\textwidth}
\includegraphics[width=\textwidth]{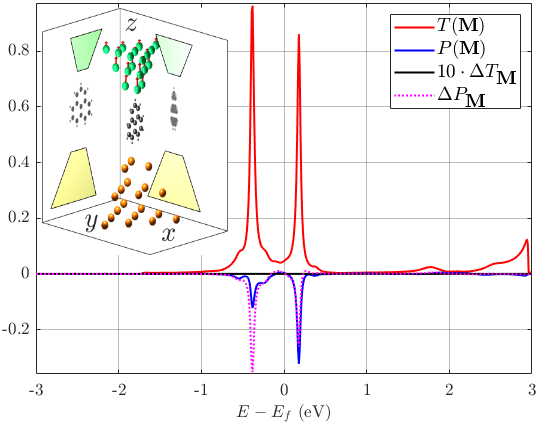}
    \caption{}
    \label{Fig2D}
\end{subfigure}
\caption{\label{Fig2} 
Dimensionless magneto-conductance ($\Delta T_{\bm{M}}=T(\bm{M})-T(-\bm{M})$) and symmetrized spin polarization ($\Delta\bm{P}_{\bm{M}}=\bm{P}(\bm{M})+\bm{P}(-\bm{M})$, component $\bm{P}\parallel\bm{M}$) in junctions with: Pb source contact with $C_{3v}$ symmetry, triangulene molecule and ferromagnetic ($\bm{M}$) Ni drain contact with $C_{3v}$ (disregarding magnetism) in different orientations of both the molecule and $\bm{M}$. Insets show the system in each subfigure, including the (ferro-)magnetization per Ni atom in red arrows whose length is illustrative. (a) Molecule aligned with the contacts, $\bm{M}=M_{y}\hat{\bm{y}}$, point group $\pazocal{G}=C_{1v}=\set{E,\sigma_{y}}$, bias $1V$. (b) Same as (a) except for $\bm{M}=M_{z}\hat{\bm{z}}$ and point group $\pazocal{G}=\set{E,\Theta\sigma_{z}}$. (c) Same as (b), with molecule rotated by 15$^{\circ}$, trivial point group $\pazocal{G}=C_{1}$. (d) Same as (c) except for no bias ($0V$). SOC is included only in the contacts, excluding the shown layers of maximum and minimum $z$. In all cases, both the SOC strength in Ni and the values of the MC have been enlarged by a factor 10 for clarity sake in the qualitative behaviour of the spin signals. No symmetries have been explicitly used in the self-consistent calculations, hence the (comparatively) small but finite $\Delta T_{\bm{M}},\Delta\bm{P}_{\bm{M}}$ in (b).} 
\end{figure}

The previous DFT calculations out of equilibrium can be expanded to compute the MC by introducing a ferromagnetic component (computational details can be found in the Methods section). Here we consider the Pb-triangulene molecular junction of Figure \ref{Fig1D}-\ref{Fig1E} with a ferromagnetic Ni drain contact that does not change the original point groups when disregarding magnetism. Other illustrative systems may be found in the Supporting Information. The results are displayed in Figure \ref{Fig2}, where the different $\bm{M}$ orientations and rotations of the molecule allow testing of the previous selection rules for the MC, in particular for the longitudinal mirror symmetry (which chiral systems lack). Note that in our case the MC is a spin-related signal stemming purely from the SOC, since in its absence (also of further spin-dependent terms other than the ferromagnetism) it would be $T^{\uparrow\uparrow}(\pm\bm{M})=T^{\downarrow\downarrow}(\mp\bm{M})$, $T^{\uparrow\downarrow}(\pm\bm{M})=T^{\downarrow\uparrow}(\mp\bm{M})=0$; so that $\Delta T_{\bm{M}}=0$. 

It can be observed in Figure \ref{Fig2A} that the MC is indeed finite, even if small compared to the conductance, for $\bm{M}$ perpendicular to the symmetry plane $\sigma_{y}$, i.e., in a transversal direction; and remarkably small (theoretically vanishing) for $\bm{M}$ along the transport direction (Figure \ref{Fig2B}) since the latter is contained in $\sigma_{y}$. The fact that the MC and $\Delta\bm{P}_{\bm{M}}$ are not strictly vanishing in this case should be explained by the fact that the mirror symmetry is not explicitly used in our self-consistent calculations. Upon rotation of the molecule, the point group of the junction can be made trivial thus enabling the emergence of MC along the (previously forbidden) transport direction, see Figure \ref{Fig2C}. Onsager's relation is verified in Figure \ref{Fig2D} for the latter asymmetric system showing that MC cannot appear in equilibrium, even in the presence of a finite spin polarization without magnetism. As observed, this result is numerically robust against the absence of symmetries, see further examples in the Supporting Information. Recalling Figure \ref{Fig1D}-\ref{Fig1E}, the MC indeed coexists with a net spin density out of equilibrium. Furthermore, these two quantities in turn coexist with a finite $\Delta\bm{P}_{\bm{M}}$, which is the smoking gun without magnetism and with or without applied voltage. 

In summary, we derive a complete set of symmetry restrictions on both spin density and MC that are valid out of equilibrium, and confirm and illustrate them via magnetic DFT quantum transport calculations, with a self-consistent treatment of the bias voltage. These selection rules emphasize and determine the important role of the geometry of the system as a whole, and help identify the underlying relations between the central quantities in spin transport phenomena, such as the CISS effect. In particular, the presence of a finite spin polarization in (or out) of equilibrium along a given direction without magnetic elements indicates that two other finite quantities will arise out of equilibrium: (1) a net spin accumulation in the system along the same direction, and (2) a MC upon introduction of a ferromagnetic detector with magnetization along the same direction.

\section{Methods}

The DFT-based quantum transport calculations have been performed with our code \texttt{ANT} \cite{palacios2001fullerene,palacios2002transport,louis2003keldysh,Jacob2011}, which is fully integrated in Gaussian09 \cite{GAUSSIAN09}. In \texttt{ANT} the electrodes outside the scattering region (not depicted in the figures) are described by a tight-binding model on a Bethe lattice, which facilitates the calculation of self-energies while keeping the symmetry of the system intact \cite{Jacob2011}. As described in Refs. \cite{Pakdel2018,DednamSAIP2022,doi:10.1021/acsnano.2c11410}, SOC is included as a post-selfconsistency first-order perturbation correction with prior optimization \cite{billy} of typically large Gaussian basis sets \cite{Zagorac2011,Laun2018,Oliviera2019,Laun2021} for the purpose of faithfully reproducing the electronic structure of the bulk electrodes. The usual Perdew-Burke-Ernzerhof (PBE) exchange-correlation functional \cite{Perdew1996} is used on account of its relatively low computational cost and ability to reproduce the electronic structure of metals reasonably well. The underestimation of the molecular gaps is of no concern here.

The equilibrium and out-of-equilibrium spin densities are obtained following \eqref{density}-\eqref{G<} employing a self-consistent Kohn-Sham Hamiltonian computed with Gaussian09 \cite{GAUSSIAN09}, imposing a convergence tolerance of $10^{-7}$ Ha. In the absence of SOC, the calculation of the equilibrium part of the spin density can be efficiently carried out by taking the integration contour along a semi circumference in the upper complex half-plane (similar to the evaluation of the charge density in the self-consistent procedure). However, upon the addition of SOC the Hamiltonian is no longer real and symmetric, hence the retarded and advanced Green's functions must be integrated separately in the upper and lower complex half-planes, respectively, as explained in Ref. \cite{Ozaki2010}. Details of the non-equilibrium implementation prior to the addition of SOC, where the integration in the bias window needs to be done along the real axis, is provided in Ref. \cite{louis2003keldysh} and this is unaltered by the addition of SOC.

In the MC calculations, the reversal of the magnetization $\bm{M}$ is achieved by swapping the spin-diagonal blocks of the fully converged unrestricted Kohn-Sham Hamiltonian, prior to the addition of SOC, once the desired degree of self-consistency has been achieved. For simplicity's sake and computational efficiency, we have considered here a minimal $sd$ basis in all Ni atoms. As stated above, the SOC strength in Ni has been consistently increased by a factor of 10, also in the Supporting Information, in order to make the MC clearly visible alongside conductance and spin polarization in a single axis system. No explicit use of symmetries has been made in any MC calculations, also in the Supporting Information


\begin{acknowledgement}
J.J.P. and M.A.G.B acknowledge financial support from Spanish MICINN (Grant Nos. PID2019-109539GB-C43 \& TED2021-131323B-I00), María de Maeztu Program for Units of Excellence in R\&D (GrantNo.CEX2018-000805-M), Comunidad Autónoma de Madrid through the Nanomag COST-CM Program (GrantNo.S2018/NMT-4321), Generalitat Valenciana through Programa Prometeo (2021/017), Centro de Computación Científica of the Universidad Autónoma de Madrid, and Red Española de Supercomputación. W.D. acknowledges usage of the High Performance Cluster infrastructure of the University of South Africa and the technical assistance of E. B. Lombardi in the attainment of a subset of the numerical results presented in this paper.
\end{acknowledgement}

The authors declare no competing financial interest.

\section{Supporting Information}

Supporting Information available:  

Same quantities as in any subfigure of Figure \ref{Fig2} (dimension-less conductance, spin polarization, magneto-conductance and sum of polarizations for opposite magnetizations) for the following systems: bare Ni-Ni contacts with different distortions or symmetries and spin directions, molecular bridge with Pt and Ni contacts and helical molecule, bare W-W-Ni contacts with three-fold rotation symmetry.  



\bibliography{SHEnano} 

\providecommand{\latin}[1]{#1}
\makeatletter
\providecommand{\doi}
  {\begingroup\let\do\@makeother\dospecials
  \catcode`\{=1 \catcode`\}=2 \doi@aux}
\providecommand{\doi@aux}[1]{\endgroup\texttt{#1}}
\makeatother
\providecommand*\mcitethebibliography{\thebibliography}
\csname @ifundefined\endcsname{endmcitethebibliography}
  {\let\endmcitethebibliography\endthebibliography}{}
\begin{mcitethebibliography}{57}
\providecommand*\natexlab[1]{#1}
\providecommand*\mciteSetBstSublistMode[1]{}
\providecommand*\mciteSetBstMaxWidthForm[2]{}
\providecommand*\mciteBstWouldAddEndPuncttrue
  {\def\EndOfBibitem{\unskip.}}
\providecommand*\mciteBstWouldAddEndPunctfalse
  {\let\EndOfBibitem\relax}
\providecommand*\mciteSetBstMidEndSepPunct[3]{}
\providecommand*\mciteSetBstSublistLabelBeginEnd[3]{}
\providecommand*\EndOfBibitem{}
\mciteSetBstSublistMode{f}
\mciteSetBstMaxWidthForm{subitem}{(\alph{mcitesubitemcount})}
\mciteSetBstSublistLabelBeginEnd
  {\mcitemaxwidthsubitemform\space}
  {\relax}
  {\relax}

\bibitem[{D'Yakonov} and {Perel'}(1971){D'Yakonov}, and
  {Perel'}]{1971JETPL..13..467D}
{D'Yakonov},~M.~I.; {Perel'},~V.~I. {Possibility of Orienting Electron Spins
  with Current}. \emph{Soviet Journal of Experimental and Theoretical Physics
  Letters} \textbf{1971}, \emph{13}, 467\relax
\mciteBstWouldAddEndPuncttrue
\mciteSetBstMidEndSepPunct{\mcitedefaultmidpunct}
{\mcitedefaultendpunct}{\mcitedefaultseppunct}\relax
\EndOfBibitem
\bibitem[Silsbee(2004)]{silsbee2004spin}
Silsbee,~R.~H. Spin--orbit induced coupling of charge current and spin
  polarization. \emph{Journal of Physics: Condensed Matter} \textbf{2004},
  \emph{16}, R179\relax
\mciteBstWouldAddEndPuncttrue
\mciteSetBstMidEndSepPunct{\mcitedefaultmidpunct}
{\mcitedefaultendpunct}{\mcitedefaultseppunct}\relax
\EndOfBibitem
\bibitem[Soumyanarayanan \latin{et~al.}(2016)Soumyanarayanan, Reyren, Fert, and
  Panagopoulos]{soumyanarayanan2016emergent}
Soumyanarayanan,~A.; Reyren,~N.; Fert,~A.; Panagopoulos,~C. Emergent phenomena
  induced by spin--orbit coupling at surfaces and interfaces. \emph{Nature}
  \textbf{2016}, \emph{539}, 509--517\relax
\mciteBstWouldAddEndPuncttrue
\mciteSetBstMidEndSepPunct{\mcitedefaultmidpunct}
{\mcitedefaultendpunct}{\mcitedefaultseppunct}\relax
\EndOfBibitem
\bibitem[Calavalle \latin{et~al.}(2022)Calavalle, Su{\'a}rez-Rodr{\'i}guez,
  Mart{\'i}n-Garc{\'i}a, Johansson, Vaz, Yang, Maznichenko, Ostanin,
  Mateo-Alonso, Chuvilin, Mertig, Gobbi, Casanova, and Hueso]{Calavalle2022}
Calavalle,~F.; Su{\'a}rez-Rodr{\'i}guez,~M.; Mart{\'i}n-Garc{\'i}a,~B.;
  Johansson,~A.; Vaz,~D.~C.; Yang,~H.; Maznichenko,~I.~V.; Ostanin,~S.;
  Mateo-Alonso,~A.; Chuvilin,~A.; Mertig,~I.; Gobbi,~M.; Casanova,~F.;
  Hueso,~L.~E. Gate-tuneable and chirality-dependent charge-to-spin conversion
  in tellurium nanowires. \emph{Nature Materials} \textbf{2022}, \emph{21},
  526--532\relax
\mciteBstWouldAddEndPuncttrue
\mciteSetBstMidEndSepPunct{\mcitedefaultmidpunct}
{\mcitedefaultendpunct}{\mcitedefaultseppunct}\relax
\EndOfBibitem
\bibitem[Sinova \latin{et~al.}(2004)Sinova, Culcer, Niu, Sinitsyn, Jungwirth,
  and MacDonald]{Sinova.PhysRevLett.92.126603}
Sinova,~J.; Culcer,~D.; Niu,~Q.; Sinitsyn,~N.~A.; Jungwirth,~T.;
  MacDonald,~A.~H. Universal Intrinsic Spin Hall Effect. \emph{Phys. Rev.
  Lett.} \textbf{2004}, \emph{92}, 126603\relax
\mciteBstWouldAddEndPuncttrue
\mciteSetBstMidEndSepPunct{\mcitedefaultmidpunct}
{\mcitedefaultendpunct}{\mcitedefaultseppunct}\relax
\EndOfBibitem
\bibitem[Valenzuela and Tinkham(2006)Valenzuela, and Tinkham]{Valenzuela2006}
Valenzuela,~S.~O.; Tinkham,~M. Direct electronic measurement of the spin Hall
  effect. \emph{Nature} \textbf{2006}, \emph{442}, 176--179\relax
\mciteBstWouldAddEndPuncttrue
\mciteSetBstMidEndSepPunct{\mcitedefaultmidpunct}
{\mcitedefaultendpunct}{\mcitedefaultseppunct}\relax
\EndOfBibitem
\bibitem[Edelstein(1990)]{EDELSTEIN1990233}
Edelstein,~V. Spin polarization of conduction electrons induced by electric
  current in two-dimensional asymmetric electron systems. \emph{Solid State
  Communications} \textbf{1990}, \emph{73}, 233--235\relax
\mciteBstWouldAddEndPuncttrue
\mciteSetBstMidEndSepPunct{\mcitedefaultmidpunct}
{\mcitedefaultendpunct}{\mcitedefaultseppunct}\relax
\EndOfBibitem
\bibitem[Ray \latin{et~al.}(1999)Ray, Ananthavel, Waldeck, and
  Naaman]{ray1999asymmetric}
Ray,~K.; Ananthavel,~S.; Waldeck,~D.; Naaman,~R. Asymmetric scattering of
  polarized electrons by organized organic films of chiral molecules.
  \emph{Science} \textbf{1999}, \emph{283}, 814--816\relax
\mciteBstWouldAddEndPuncttrue
\mciteSetBstMidEndSepPunct{\mcitedefaultmidpunct}
{\mcitedefaultendpunct}{\mcitedefaultseppunct}\relax
\EndOfBibitem
\bibitem[G{\"o}hler \latin{et~al.}(2011)G{\"o}hler, Hamelbeck, Markus, Kettner,
  Hanne, Vager, Naaman, and Zacharias]{gohler2011spin}
G{\"o}hler,~B.; Hamelbeck,~V.; Markus,~T.; Kettner,~M.; Hanne,~G.; Vager,~Z.;
  Naaman,~R.; Zacharias,~H. Spin selectivity in electron transmission through
  self-assembled monolayers of double-stranded DNA. \emph{Science}
  \textbf{2011}, \emph{331}, 894--897\relax
\mciteBstWouldAddEndPuncttrue
\mciteSetBstMidEndSepPunct{\mcitedefaultmidpunct}
{\mcitedefaultendpunct}{\mcitedefaultseppunct}\relax
\EndOfBibitem
\bibitem[Naaman and Waldeck(2012)Naaman, and Waldeck]{Naaman2012}
Naaman,~R.; Waldeck,~D.~H. {Chiral-Induced Spin Selectivity Effect}.
  \emph{Journal of Physical Chemistry C} \textbf{2012}, \emph{3},
  2178--2187\relax
\mciteBstWouldAddEndPuncttrue
\mciteSetBstMidEndSepPunct{\mcitedefaultmidpunct}
{\mcitedefaultendpunct}{\mcitedefaultseppunct}\relax
\EndOfBibitem
\bibitem[Naaman \latin{et~al.}(2019)Naaman, Paltiel, and Waldeck]{Naaman2019}
Naaman,~R.; Paltiel,~Y.; Waldeck,~D.~H. {Chiral molecules and the electron
  spin}. \emph{Nature Reviews Chemistry} \textbf{2019}, \emph{3},
  250--260\relax
\mciteBstWouldAddEndPuncttrue
\mciteSetBstMidEndSepPunct{\mcitedefaultmidpunct}
{\mcitedefaultendpunct}{\mcitedefaultseppunct}\relax
\EndOfBibitem
\bibitem[Naaman \latin{et~al.}(2020)Naaman, Paltiel, and
  Waldeck]{naaman2020chiral}
Naaman,~R.; Paltiel,~Y.; Waldeck,~D.~H. Chiral molecules and the spin
  selectivity effect. \emph{The Journal of Physical Chemistry Letters}
  \textbf{2020}, \emph{11}, 3660--3666\relax
\mciteBstWouldAddEndPuncttrue
\mciteSetBstMidEndSepPunct{\mcitedefaultmidpunct}
{\mcitedefaultendpunct}{\mcitedefaultseppunct}\relax
\EndOfBibitem
\bibitem[Chiesa \latin{et~al.}(2023)Chiesa, Privitera, Macaluso, Mannini,
  Bittl, Naaman, Wasielewski, Sessoli, and Carretta]{chiesa2023chirality}
Chiesa,~A.; Privitera,~A.; Macaluso,~E.; Mannini,~M.; Bittl,~R.; Naaman,~R.;
  Wasielewski,~M.~R.; Sessoli,~R.; Carretta,~S. Chirality-Induced Spin
  Selectivity: An Enabling Technology for Quantum Applications. \emph{Advanced
  Materials} \textbf{2023}, 2300472\relax
\mciteBstWouldAddEndPuncttrue
\mciteSetBstMidEndSepPunct{\mcitedefaultmidpunct}
{\mcitedefaultendpunct}{\mcitedefaultseppunct}\relax
\EndOfBibitem
\bibitem[Aiello \latin{et~al.}(2022)Aiello, Abendroth, Abbas, Afanasev,
  Agarwal, Banerjee, Beratan, Belling, Berche, Botana, Caram, Celardo,
  Cuniberti, Garcia-Etxarri, Dianat, Diez-Perez, Guo, Gutierrez, Herrmann,
  Hihath, Kale, Kurian, Lai, Liu, Lopez, Medina, Mujica, Naaman, Noormandipour,
  Palma, Paltiel, Petuskey, Ribeiro-Silva, Saenz, Santos, Solyanik-Gorgone,
  Sorger, Stemer, Ugalde, Valdes-Curiel, Varela, Waldeck, Wasielewski, Weiss,
  Zacharias, and Wang]{Aiello2022}
Aiello,~C.~D. \latin{et~al.}  A Chirality-Based Quantum Leap. \emph{ACS Nano}
  \textbf{2022}, \emph{16}, 4989--5035, PMID: 35318848\relax
\mciteBstWouldAddEndPuncttrue
\mciteSetBstMidEndSepPunct{\mcitedefaultmidpunct}
{\mcitedefaultendpunct}{\mcitedefaultseppunct}\relax
\EndOfBibitem
\bibitem[Naaman \latin{et~al.}(2020)Naaman, Paltiel, and
  Waldeck]{naaman2020chiralB}
Naaman,~R.; Paltiel,~Y.; Waldeck,~D.~H. Chiral induced spin selectivity gives a
  new twist on spin-control in chemistry. \emph{Accounts of Chemical Research}
  \textbf{2020}, \emph{53}, 2659--2667\relax
\mciteBstWouldAddEndPuncttrue
\mciteSetBstMidEndSepPunct{\mcitedefaultmidpunct}
{\mcitedefaultendpunct}{\mcitedefaultseppunct}\relax
\EndOfBibitem
\bibitem[Forment-Aliaga and Gaita-Ari{\~n}o(2022)Forment-Aliaga, and
  Gaita-Ari{\~n}o]{forment2022chiral}
Forment-Aliaga,~A.; Gaita-Ari{\~n}o,~A. Chiral, magnetic, molecule-based
  materials: A chemical path toward spintronics and quantum nanodevices.
  \emph{Journal of Applied Physics} \textbf{2022}, \emph{132}, 180901\relax
\mciteBstWouldAddEndPuncttrue
\mciteSetBstMidEndSepPunct{\mcitedefaultmidpunct}
{\mcitedefaultendpunct}{\mcitedefaultseppunct}\relax
\EndOfBibitem
\bibitem[Liu \latin{et~al.}(2020)Liu, Wang, Wang, Shi, Gao, Feng, Deng, Hu,
  Lochner, Schlottmann, von Molnár, Li, Zhao, and Xiong]{acsnano.0c07438}
Liu,~T.; Wang,~X.; Wang,~H.; Shi,~G.; Gao,~F.; Feng,~H.; Deng,~H.; Hu,~L.;
  Lochner,~E.; Schlottmann,~P.; von Molnár,~S.; Li,~Y.; Zhao,~J.; Xiong,~P.
  Linear and Nonlinear Two-Terminal Spin-Valve Effect from Chirality-Induced
  Spin Selectivity. \emph{ACS Nano} \textbf{2020}, \emph{14},
  15983--15991\relax
\mciteBstWouldAddEndPuncttrue
\mciteSetBstMidEndSepPunct{\mcitedefaultmidpunct}
{\mcitedefaultendpunct}{\mcitedefaultseppunct}\relax
\EndOfBibitem
\bibitem[Maslyuk \latin{et~al.}(2018)Maslyuk, Gutierrez, Dianat, Mujica, and
  Cuniberti]{Maslyuk2018}
Maslyuk,~V.~V.; Gutierrez,~R.; Dianat,~A.; Mujica,~V.; Cuniberti,~G. Enhanced
  Magnetoresistance in Chiral Molecular Junctions. \emph{The Journal of
  Physical Chemistry Letters} \textbf{2018}, \emph{9}, 5453--5459, PMID:
  30188726\relax
\mciteBstWouldAddEndPuncttrue
\mciteSetBstMidEndSepPunct{\mcitedefaultmidpunct}
{\mcitedefaultendpunct}{\mcitedefaultseppunct}\relax
\EndOfBibitem
\bibitem[Fransson(2022)]{fransson2022chiral}
Fransson,~J. The Chiral Induced Spin Selectivity Effect What It Is, What It Is
  Not, And Why It Matters. \emph{Israel Journal of Chemistry} \textbf{2022},
  e202200046\relax
\mciteBstWouldAddEndPuncttrue
\mciteSetBstMidEndSepPunct{\mcitedefaultmidpunct}
{\mcitedefaultendpunct}{\mcitedefaultseppunct}\relax
\EndOfBibitem
\bibitem[Gersten \latin{et~al.}(2013)Gersten, Kaasbjerg, and
  Nitzan]{gersten2013induced}
Gersten,~J.; Kaasbjerg,~K.; Nitzan,~A. Induced spin filtering in electron
  transmission through chiral molecular layers adsorbed on metals with strong
  spin-orbit coupling. \emph{The Journal of chemical physics} \textbf{2013},
  \emph{139}, 114111\relax
\mciteBstWouldAddEndPuncttrue
\mciteSetBstMidEndSepPunct{\mcitedefaultmidpunct}
{\mcitedefaultendpunct}{\mcitedefaultseppunct}\relax
\EndOfBibitem
\bibitem[Zöllner \latin{et~al.}(2020)Zöllner, Varela, Medina, Mujica, and
  Herrmann]{Zollner2020a}
Zöllner,~M.~S.; Varela,~S.; Medina,~E.; Mujica,~V.; Herrmann,~C. Insight into
  the Origin of Chiral-Induced Spin Selectivity from a Symmetry Analysis of
  Electronic Transmission. \emph{Journal of Chemical Theory and Computation}
  \textbf{2020}, \emph{16}, 2914--2929\relax
\mciteBstWouldAddEndPuncttrue
\mciteSetBstMidEndSepPunct{\mcitedefaultmidpunct}
{\mcitedefaultendpunct}{\mcitedefaultseppunct}\relax
\EndOfBibitem
\bibitem[Zöllner \latin{et~al.}(2020)Zöllner, Saghatchi, Mujica, and
  Herrmann]{Zollner2020b}
Zöllner,~M.~S.; Saghatchi,~A.; Mujica,~V.; Herrmann,~C. Influence of
  Electronic Structure Modeling and Junction Structure on First-Principles
  Chiral Induced Spin Selectivity. \emph{Journal of Chemical Theory and
  Computation} \textbf{2020}, \emph{16}, 7357--7371\relax
\mciteBstWouldAddEndPuncttrue
\mciteSetBstMidEndSepPunct{\mcitedefaultmidpunct}
{\mcitedefaultendpunct}{\mcitedefaultseppunct}\relax
\EndOfBibitem
\bibitem[Alwan and Dubi(2021)Alwan, and Dubi]{alwan2021spinterface}
Alwan,~S.; Dubi,~Y. Spinterface Origin for the Chirality-Induced
  Spin-Selectivity Effect. \emph{Journal of the American Chemical Society}
  \textbf{2021}, \emph{143}, 14235--14241\relax
\mciteBstWouldAddEndPuncttrue
\mciteSetBstMidEndSepPunct{\mcitedefaultmidpunct}
{\mcitedefaultendpunct}{\mcitedefaultseppunct}\relax
\EndOfBibitem
\bibitem[Liu \latin{et~al.}(2021)Liu, Xiao, Koo, and Yan]{Liu2021}
Liu,~Y.; Xiao,~J.; Koo,~J.; Yan,~B. {Chirality-driven topological electronic
  structure of DNA-like materials}. \emph{Nature Materials} \textbf{2021},
  \emph{20}, 638--644\relax
\mciteBstWouldAddEndPuncttrue
\mciteSetBstMidEndSepPunct{\mcitedefaultmidpunct}
{\mcitedefaultendpunct}{\mcitedefaultseppunct}\relax
\EndOfBibitem
\bibitem[Dednam \latin{et~al.}(2023)Dednam, García-Blázquez, Zotti, Lombardi,
  Sabater, Pakdel, and Palacios]{doi:10.1021/acsnano.2c11410}
Dednam,~W.; García-Blázquez,~M.~A.; Zotti,~L.~A.; Lombardi,~E.~B.;
  Sabater,~C.; Pakdel,~S.; Palacios,~J.~J. A Group-Theoretic Approach to the
  Origin of Chirality-Induced Spin-Selectivity in Nonmagnetic Molecular
  Junctions. \emph{ACS Nano} \textbf{2023}, \emph{17}, 6452--6465, PMID:
  36947721\relax
\mciteBstWouldAddEndPuncttrue
\mciteSetBstMidEndSepPunct{\mcitedefaultmidpunct}
{\mcitedefaultendpunct}{\mcitedefaultseppunct}\relax
\EndOfBibitem
\bibitem[Guo \latin{et~al.}(2016)Guo, Pan, Fang, Xie, and Sun]{Guo2016}
Guo,~A.~M.; Pan,~T.~R.; Fang,~T.~F.; Xie,~X.~C.; Sun,~Q.~F. {Spin selectivity
  effect in achiral molecular systems}. \emph{Physical Review B} \textbf{2016},
  \emph{94}, 1--5\relax
\mciteBstWouldAddEndPuncttrue
\mciteSetBstMidEndSepPunct{\mcitedefaultmidpunct}
{\mcitedefaultendpunct}{\mcitedefaultseppunct}\relax
\EndOfBibitem
\bibitem[Wolf \latin{et~al.}(2022)Wolf, Liu, Xiao, Park, and
  Yan]{acsnano.2c07088}
Wolf,~Y.; Liu,~Y.; Xiao,~J.; Park,~N.; Yan,~B. Unusual Spin Polarization in the
  Chirality-Induced Spin Selectivity. \emph{ACS Nano} \textbf{2022}, \emph{0},
  null\relax
\mciteBstWouldAddEndPuncttrue
\mciteSetBstMidEndSepPunct{\mcitedefaultmidpunct}
{\mcitedefaultendpunct}{\mcitedefaultseppunct}\relax
\EndOfBibitem
\bibitem[Yang \latin{et~al.}(2019)Yang, van~der Wal, and van
  Wees]{yang2019spin}
Yang,~X.; van~der Wal,~C.~H.; van Wees,~B.~J. Spin-dependent electron
  transmission model for chiral molecules in mesoscopic devices. \emph{Physical
  Review B} \textbf{2019}, \emph{99}, 024418\relax
\mciteBstWouldAddEndPuncttrue
\mciteSetBstMidEndSepPunct{\mcitedefaultmidpunct}
{\mcitedefaultendpunct}{\mcitedefaultseppunct}\relax
\EndOfBibitem
\bibitem[Yang \latin{et~al.}(2020)Yang, van~der Wal, and van
  Wees]{yang2020detecting}
Yang,~X.; van~der Wal,~C.~H.; van Wees,~B.~J. Detecting chirality in
  two-terminal electronic nanodevices. \emph{Nano letters} \textbf{2020},
  \emph{20}, 6148--6154\relax
\mciteBstWouldAddEndPuncttrue
\mciteSetBstMidEndSepPunct{\mcitedefaultmidpunct}
{\mcitedefaultendpunct}{\mcitedefaultseppunct}\relax
\EndOfBibitem
\bibitem[Buttiker(1988)]{buttiker1988symmetry}
Buttiker,~M. Symmetry of electrical conduction. \emph{IBM Journal of Research
  and Development} \textbf{1988}, \emph{32}, 317--334\relax
\mciteBstWouldAddEndPuncttrue
\mciteSetBstMidEndSepPunct{\mcitedefaultmidpunct}
{\mcitedefaultendpunct}{\mcitedefaultseppunct}\relax
\EndOfBibitem
\bibitem[Zhai and Xu(2005)Zhai, and Xu]{zhai2005symmetry}
Zhai,~F.; Xu,~H. Symmetry of spin transport in two-terminal waveguides with a
  spin-orbital interaction and magnetic field modulations. \emph{Physical
  review letters} \textbf{2005}, \emph{94}, 246601\relax
\mciteBstWouldAddEndPuncttrue
\mciteSetBstMidEndSepPunct{\mcitedefaultmidpunct}
{\mcitedefaultendpunct}{\mcitedefaultseppunct}\relax
\EndOfBibitem
\bibitem[Naskar \latin{et~al.}(0)Naskar, Mujica, and Herrmann]{naskar2022}
Naskar,~S.; Mujica,~V.; Herrmann,~C. Chiral-Induced Spin Selectivity and
  Non-equilibrium Spin Accumulation in Molecules and Interfaces: A
  First-Principles Study. \emph{The Journal of Physical Chemistry Letters}
  \textbf{0}, \emph{0}, 694--701, PMID: 36638217\relax
\mciteBstWouldAddEndPuncttrue
\mciteSetBstMidEndSepPunct{\mcitedefaultmidpunct}
{\mcitedefaultendpunct}{\mcitedefaultseppunct}\relax
\EndOfBibitem
\bibitem[Fransson(2019)]{acs.jpclett.9b02929}
Fransson,~J. Chirality-Induced Spin Selectivity: The Role of Electron
  Correlations. \emph{The Journal of Physical Chemistry Letters} \textbf{2019},
  \emph{10}, 7126--7132\relax
\mciteBstWouldAddEndPuncttrue
\mciteSetBstMidEndSepPunct{\mcitedefaultmidpunct}
{\mcitedefaultendpunct}{\mcitedefaultseppunct}\relax
\EndOfBibitem
\bibitem[Fransson(2021)]{acs.nanolett.1c00183}
Fransson,~J. Charge Redistribution and Spin Polarization Driven by Correlation
  Induced Electron Exchange in Chiral Molecules. \emph{Nano Letters}
  \textbf{2021}, \emph{21}, 3026--3032\relax
\mciteBstWouldAddEndPuncttrue
\mciteSetBstMidEndSepPunct{\mcitedefaultmidpunct}
{\mcitedefaultendpunct}{\mcitedefaultseppunct}\relax
\EndOfBibitem
\bibitem[Du \latin{et~al.}(2020)Du, Fu, and Wu]{PhysRevB.102.035431}
Du,~G.-F.; Fu,~H.-H.; Wu,~R. Vibration-enhanced spin-selective transport of
  electrons in the DNA double helix. \emph{Phys. Rev. B} \textbf{2020},
  \emph{102}, 035431\relax
\mciteBstWouldAddEndPuncttrue
\mciteSetBstMidEndSepPunct{\mcitedefaultmidpunct}
{\mcitedefaultendpunct}{\mcitedefaultseppunct}\relax
\EndOfBibitem
\bibitem[Fransson(2020)]{PhysRevB.102.235416}
Fransson,~J. Vibrational origin of exchange splitting and ''chiral-induced spin
  selectivity. \emph{Phys. Rev. B} \textbf{2020}, \emph{102}, 235416\relax
\mciteBstWouldAddEndPuncttrue
\mciteSetBstMidEndSepPunct{\mcitedefaultmidpunct}
{\mcitedefaultendpunct}{\mcitedefaultseppunct}\relax
\EndOfBibitem
\bibitem[Zhang \latin{et~al.}(2020)Zhang, Hao, Qin, Xie, and
  Qu]{PhysRevB.102.214303}
Zhang,~L.; Hao,~Y.; Qin,~W.; Xie,~S.; Qu,~F. Chiral-induced spin selectivity: A
  polaron transport model. \emph{Phys. Rev. B} \textbf{2020}, \emph{102},
  214303\relax
\mciteBstWouldAddEndPuncttrue
\mciteSetBstMidEndSepPunct{\mcitedefaultmidpunct}
{\mcitedefaultendpunct}{\mcitedefaultseppunct}\relax
\EndOfBibitem
\bibitem[Dubi(2022)]{dubi2022spinterface}
Dubi,~Y. Spinterface chirality-induced spin selectivity effect in
  bio-molecules. \emph{Chemical science} \textbf{2022}, \emph{13},
  10878--10883\relax
\mciteBstWouldAddEndPuncttrue
\mciteSetBstMidEndSepPunct{\mcitedefaultmidpunct}
{\mcitedefaultendpunct}{\mcitedefaultseppunct}\relax
\EndOfBibitem
\bibitem[Soriano and Palacios(2014)Soriano, and Palacios]{soriano2014theory}
Soriano,~M.; Palacios,~J. Theory of projections with nonorthogonal basis sets:
  Partitioning techniques and effective Hamiltonians. \emph{Physical Review B}
  \textbf{2014}, \emph{90}, 075128\relax
\mciteBstWouldAddEndPuncttrue
\mciteSetBstMidEndSepPunct{\mcitedefaultmidpunct}
{\mcitedefaultendpunct}{\mcitedefaultseppunct}\relax
\EndOfBibitem
\bibitem[Szabo and Ostlund(2012)Szabo, and Ostlund]{szabo2012modern}
Szabo,~A.; Ostlund,~N.~S. \emph{Modern quantum chemistry: introduction to
  advanced electronic structure theory}; Courier Corporation, 2012\relax
\mciteBstWouldAddEndPuncttrue
\mciteSetBstMidEndSepPunct{\mcitedefaultmidpunct}
{\mcitedefaultendpunct}{\mcitedefaultseppunct}\relax
\EndOfBibitem
\bibitem[Taylor \latin{et~al.}(2001)Taylor, Guo, and Wang]{taylor2001ab}
Taylor,~J.; Guo,~H.; Wang,~J. Ab initio modeling of quantum transport
  properties of molecular electronic devices. \emph{Physical Review B}
  \textbf{2001}, \emph{63}, 245407\relax
\mciteBstWouldAddEndPuncttrue
\mciteSetBstMidEndSepPunct{\mcitedefaultmidpunct}
{\mcitedefaultendpunct}{\mcitedefaultseppunct}\relax
\EndOfBibitem
\bibitem[Jacob and Palacios(2011)Jacob, and Palacios]{Jacob2011}
Jacob,~D.; Palacios,~J.~J. Critical comparison of electrode models in density
  functional theory based quantum transport calculations. \emph{The Journal of
  Chemical Physics} \textbf{2011}, \emph{134}, 044118\relax
\mciteBstWouldAddEndPuncttrue
\mciteSetBstMidEndSepPunct{\mcitedefaultmidpunct}
{\mcitedefaultendpunct}{\mcitedefaultseppunct}\relax
\EndOfBibitem
\bibitem[Nikoli{\'c} and Souma(2005)Nikoli{\'c}, and
  Souma]{nikolic2005decoherence}
Nikoli{\'c},~B.~K.; Souma,~S. Decoherence of transported spin in multichannel
  spin-orbit-coupled spintronic devices: Scattering approach to spin-density
  matrix from the ballistic to the localized regime. \emph{Physical Review B}
  \textbf{2005}, \emph{71}, 195328\relax
\mciteBstWouldAddEndPuncttrue
\mciteSetBstMidEndSepPunct{\mcitedefaultmidpunct}
{\mcitedefaultendpunct}{\mcitedefaultseppunct}\relax
\EndOfBibitem
\bibitem[Palacios \latin{et~al.}(2001)Palacios, P\'erez-Jim\'enez, Louis, and
  Verg\'es]{palacios2001fullerene}
Palacios,~J.~J.; P\'erez-Jim\'enez,~A.~J.; Louis,~E.; Verg\'es,~J.~A.
  Fullerene-based molecular nanobridges: A first-principles study. \emph{Phys.
  Rev. B} \textbf{2001}, \emph{64}, 115411\relax
\mciteBstWouldAddEndPuncttrue
\mciteSetBstMidEndSepPunct{\mcitedefaultmidpunct}
{\mcitedefaultendpunct}{\mcitedefaultseppunct}\relax
\EndOfBibitem
\bibitem[Palacios \latin{et~al.}(2002)Palacios, P\'erez-Jim\'enez, Louis,
  SanFabi{\'a}n, and Verg\'es]{palacios2002transport}
Palacios,~J.~J.; P\'erez-Jim\'enez,~A.~J.; Louis,~E.; SanFabi{\'a}n,~E.;
  Verg\'es,~J.~A. First-principles approach to electrical transport in
  atomic-scale nanostructures. \emph{Phys. Rev. B} \textbf{2002}, \emph{66},
  035322\relax
\mciteBstWouldAddEndPuncttrue
\mciteSetBstMidEndSepPunct{\mcitedefaultmidpunct}
{\mcitedefaultendpunct}{\mcitedefaultseppunct}\relax
\EndOfBibitem
\bibitem[Louis \latin{et~al.}(2003)Louis, Verg\'es, Palacios,
  P\'erez-Jim\'enez, and SanFabi{\'a}n]{louis2003keldysh}
Louis,~E.; Verg\'es,~J.~A.; Palacios,~J.~J.; P\'erez-Jim\'enez,~A.~J.;
  SanFabi{\'a}n,~E. Implementing the Keldysh formalism into ab initio methods
  for the calculation of quantum transport: Application to metallic
  nanocontacts. \emph{Phys. Rev. B} \textbf{2003}, \emph{67}, 155321\relax
\mciteBstWouldAddEndPuncttrue
\mciteSetBstMidEndSepPunct{\mcitedefaultmidpunct}
{\mcitedefaultendpunct}{\mcitedefaultseppunct}\relax
\EndOfBibitem
\bibitem[Frisch \latin{et~al.}()Frisch, \latin{et~al.} others]{GAUSSIAN09}
Frisch,~M.~J., \latin{et~al.}  Computer code {GAUSSIAN09}, {R}evision {C.01},
  {G}aussian, {I}nc. {W}allingford, {CT}, 2009\relax
\mciteBstWouldAddEndPuncttrue
\mciteSetBstMidEndSepPunct{\mcitedefaultmidpunct}
{\mcitedefaultendpunct}{\mcitedefaultseppunct}\relax
\EndOfBibitem
\bibitem[Pakdel \latin{et~al.}(2018)Pakdel, Pourfath, and Palacios]{Pakdel2018}
Pakdel,~S.; Pourfath,~M.; Palacios,~J.~J. An implementation of spin–orbit
  coupling for band structure calculations with Gaussian basis sets:
  Two-dimensional topological crystals of Sb and Bi. \emph{Beilstein J.
  Nanotechnol.} \textbf{2018}, \emph{9}, 1015\relax
\mciteBstWouldAddEndPuncttrue
\mciteSetBstMidEndSepPunct{\mcitedefaultmidpunct}
{\mcitedefaultendpunct}{\mcitedefaultseppunct}\relax
\EndOfBibitem
\bibitem[Dednam \latin{et~al.}(2022)Dednam, Zotti, Pakdel, Lombardi, and
  Palacios]{DednamSAIP2022}
Dednam,~W.; Zotti,~L.~A.; Pakdel,~S.; Lombardi,~E.; Palacios,~J.
  Spin-imbalances in non-magnetic nano-systems: Using non-equilibrium Green’s
  function DFT to model spin-selective phenomena mediated by spin-orbit
  coupling. \emph{Proc. 65th Annual Conf. South African Institute of Physics}
  \textbf{2022}, \emph{65}, 25--30\relax
\mciteBstWouldAddEndPuncttrue
\mciteSetBstMidEndSepPunct{\mcitedefaultmidpunct}
{\mcitedefaultendpunct}{\mcitedefaultseppunct}\relax
\EndOfBibitem
\bibitem[Towler()]{billy}
Towler,~M. Computer code \texttt{billy} for optimizing \texttt{CRYSTAL14} basis
  sets. Available from \url{https://vallico.net/mike_towler/crystal.html}\relax
\mciteBstWouldAddEndPuncttrue
\mciteSetBstMidEndSepPunct{\mcitedefaultmidpunct}
{\mcitedefaultendpunct}{\mcitedefaultseppunct}\relax
\EndOfBibitem
\bibitem[Zagorac \latin{et~al.}(2011)Zagorac, Doll, Sch{\"{o}}n, and
  Jansen]{Zagorac2011}
Zagorac,~D.; Doll,~K.; Sch{\"{o}}n,~J.~C.; Jansen,~M. {Ab initio structure
  prediction for lead sulfide at standard and elevated pressures}.
  \emph{Physical Review B - Condensed Matter and Materials Physics}
  \textbf{2011}, \emph{84}, 1--13\relax
\mciteBstWouldAddEndPuncttrue
\mciteSetBstMidEndSepPunct{\mcitedefaultmidpunct}
{\mcitedefaultendpunct}{\mcitedefaultseppunct}\relax
\EndOfBibitem
\bibitem[Laun \latin{et~al.}(2018)Laun, Vilela~Oliveira, and Bredow]{Laun2018}
Laun,~J.; Vilela~Oliveira,~D.; Bredow,~T. Consistent gaussian basis sets of
  double- and triple-zeta valence with polarization quality of the fifth period
  for solid-state calculations. \emph{Journal of Computational Chemistry}
  \textbf{2018}, \emph{39}, 1285--1290\relax
\mciteBstWouldAddEndPuncttrue
\mciteSetBstMidEndSepPunct{\mcitedefaultmidpunct}
{\mcitedefaultendpunct}{\mcitedefaultseppunct}\relax
\EndOfBibitem
\bibitem[Vilela~Oliveira \latin{et~al.}(2019)Vilela~Oliveira, Laun, Peintinger,
  and Bredow]{Oliviera2019}
Vilela~Oliveira,~D.; Laun,~J.; Peintinger,~M.~F.; Bredow,~T. BSSE-correction
  scheme for consistent gaussian basis sets of double- and triple-zeta valence
  with polarization quality for solid-state calculations. \emph{Journal of
  Computational Chemistry} \textbf{2019}, \emph{40}, 2364--2376\relax
\mciteBstWouldAddEndPuncttrue
\mciteSetBstMidEndSepPunct{\mcitedefaultmidpunct}
{\mcitedefaultendpunct}{\mcitedefaultseppunct}\relax
\EndOfBibitem
\bibitem[Laun and Bredow(2021)Laun, and Bredow]{Laun2021}
Laun,~J.; Bredow,~T. BSSE-corrected consistent Gaussian basis sets of
  triple-zeta valence with polarization quality of the sixth period for
  solid-state calculations. \emph{Journal of Computational Chemistry}
  \textbf{2021}, \emph{42}, 1064--1072\relax
\mciteBstWouldAddEndPuncttrue
\mciteSetBstMidEndSepPunct{\mcitedefaultmidpunct}
{\mcitedefaultendpunct}{\mcitedefaultseppunct}\relax
\EndOfBibitem
\bibitem[Perdew \latin{et~al.}(1996)Perdew, Burke, and Ernzerhof]{Perdew1996}
Perdew,~J.~P.; Burke,~K.; Ernzerhof,~M. {Generalized gradient approximation
  made simple}. \emph{Physical Review Letters} \textbf{1996}, \emph{77},
  3865--3868\relax
\mciteBstWouldAddEndPuncttrue
\mciteSetBstMidEndSepPunct{\mcitedefaultmidpunct}
{\mcitedefaultendpunct}{\mcitedefaultseppunct}\relax
\EndOfBibitem
\bibitem[Ozaki \latin{et~al.}(2010)Ozaki, Nishio, and Kino]{Ozaki2010}
Ozaki,~T.; Nishio,~K.; Kino,~H. {Efficient implementation of the nonequilibrium
  Green function method for electronic transport calculations}. \emph{Phys.
  Rev. B} \textbf{2010}, \emph{81}, 035116\relax
\mciteBstWouldAddEndPuncttrue
\mciteSetBstMidEndSepPunct{\mcitedefaultmidpunct}
{\mcitedefaultendpunct}{\mcitedefaultseppunct}\relax
\EndOfBibitem
\end{mcitethebibliography}

\end{document}









\section{Further MC results}

\subsection{Ni($C_{4}$)-Ni($C_{4}$)}

\begin{figure}[h!]
\centering 
\includegraphics[width=0.5\textwidth]{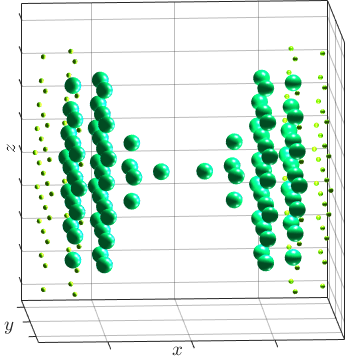}
\caption{Reference structure with point group $\pazocal{G}=D_{4h}$ ($4/mmm$)}
\end{figure}

\begin{figure}[h!]
\centering 
\includegraphics[width=0.45\textwidth]{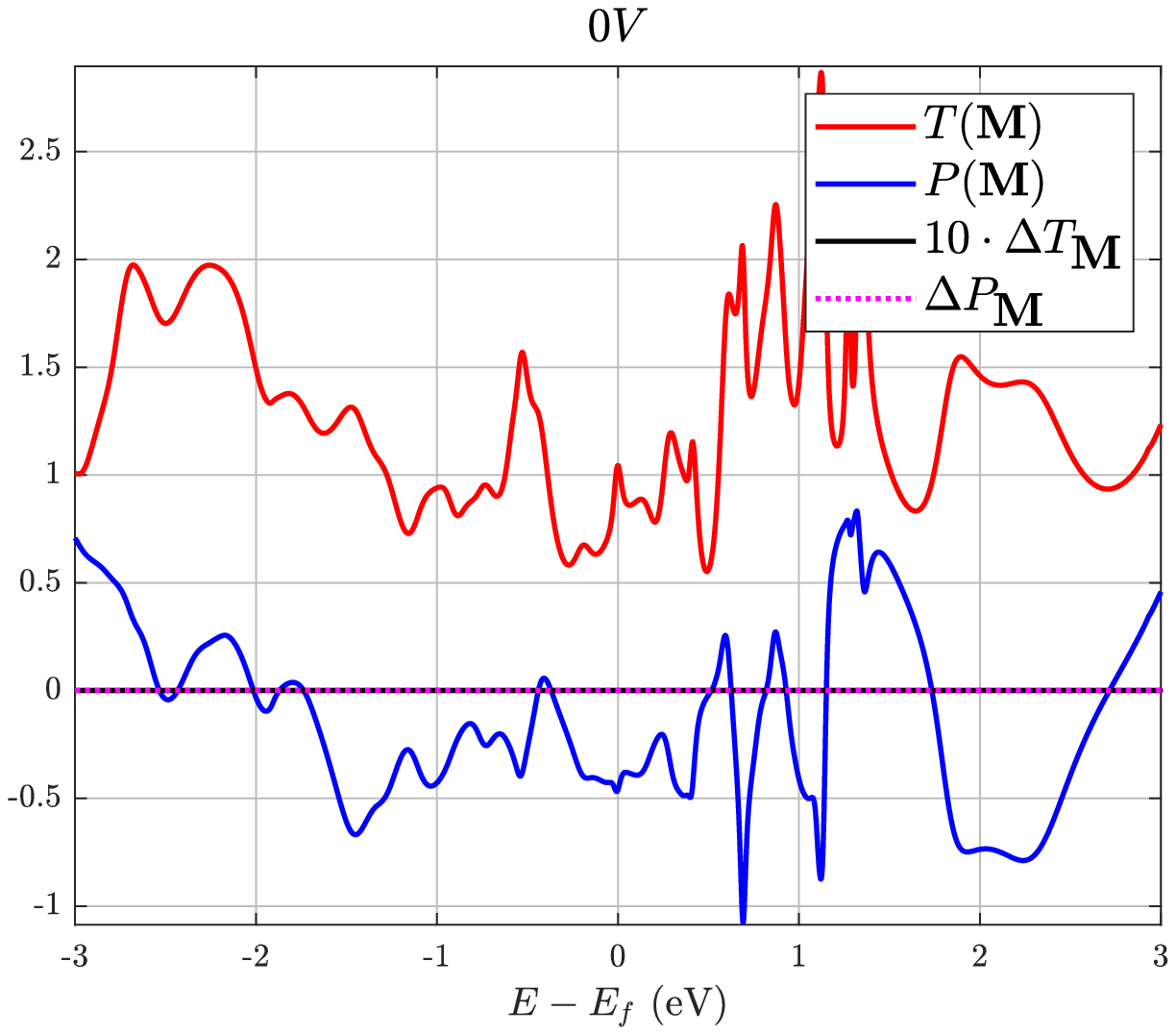}
\includegraphics[width=0.45\textwidth]{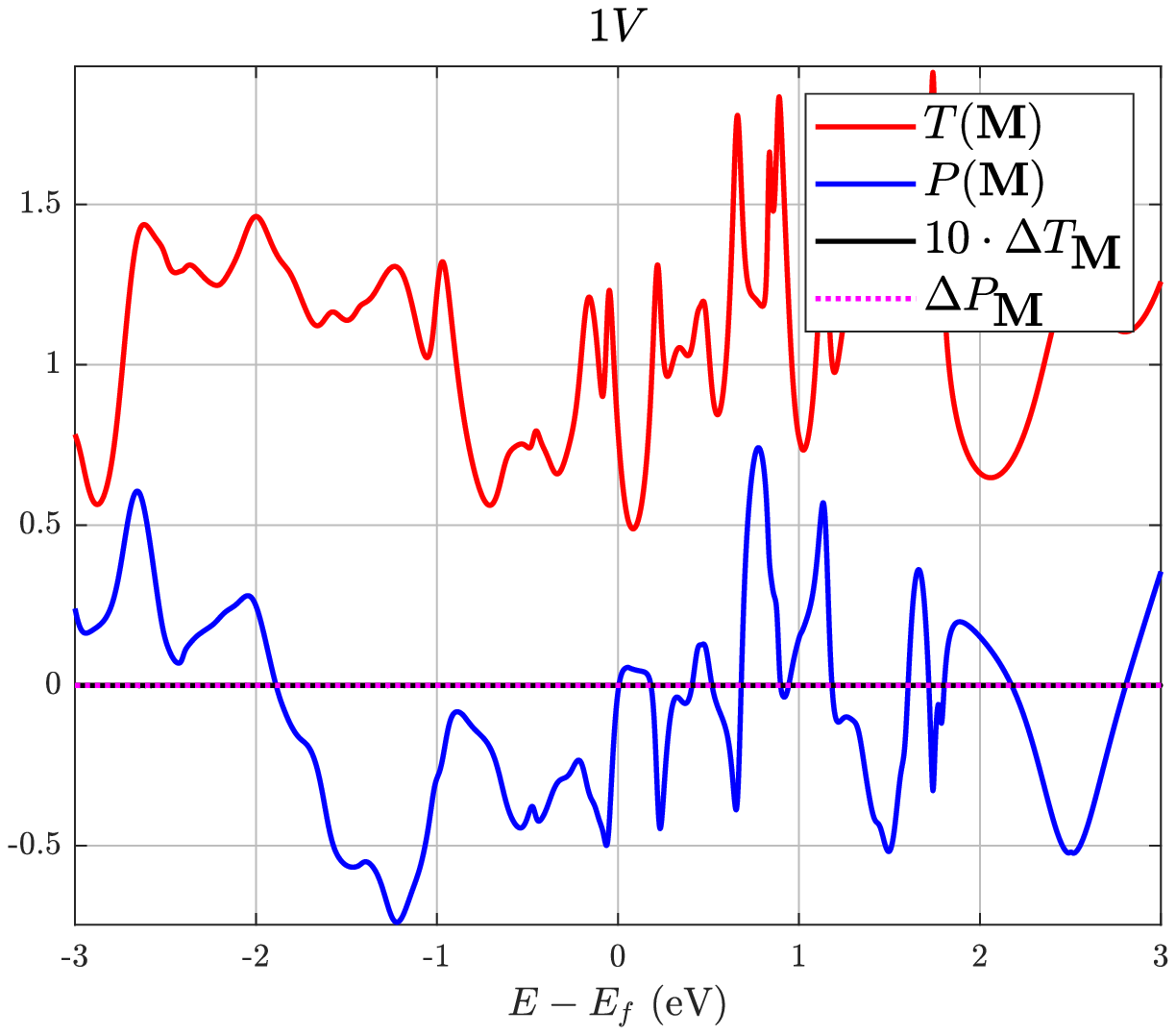}
\caption{
$\bm{M}=M\hat{\bm{z}}$ (transversal). Full symmetry: $\pazocal{G}=\left\{\begin{aligned}D_{4h}\text{, without magnetism}&\\m'm'm\text{, with magnetism}&\end{aligned}\right\}$}
\end{figure}

\begin{figure}[h!]
\centering 
\includegraphics[width=0.45\textwidth]{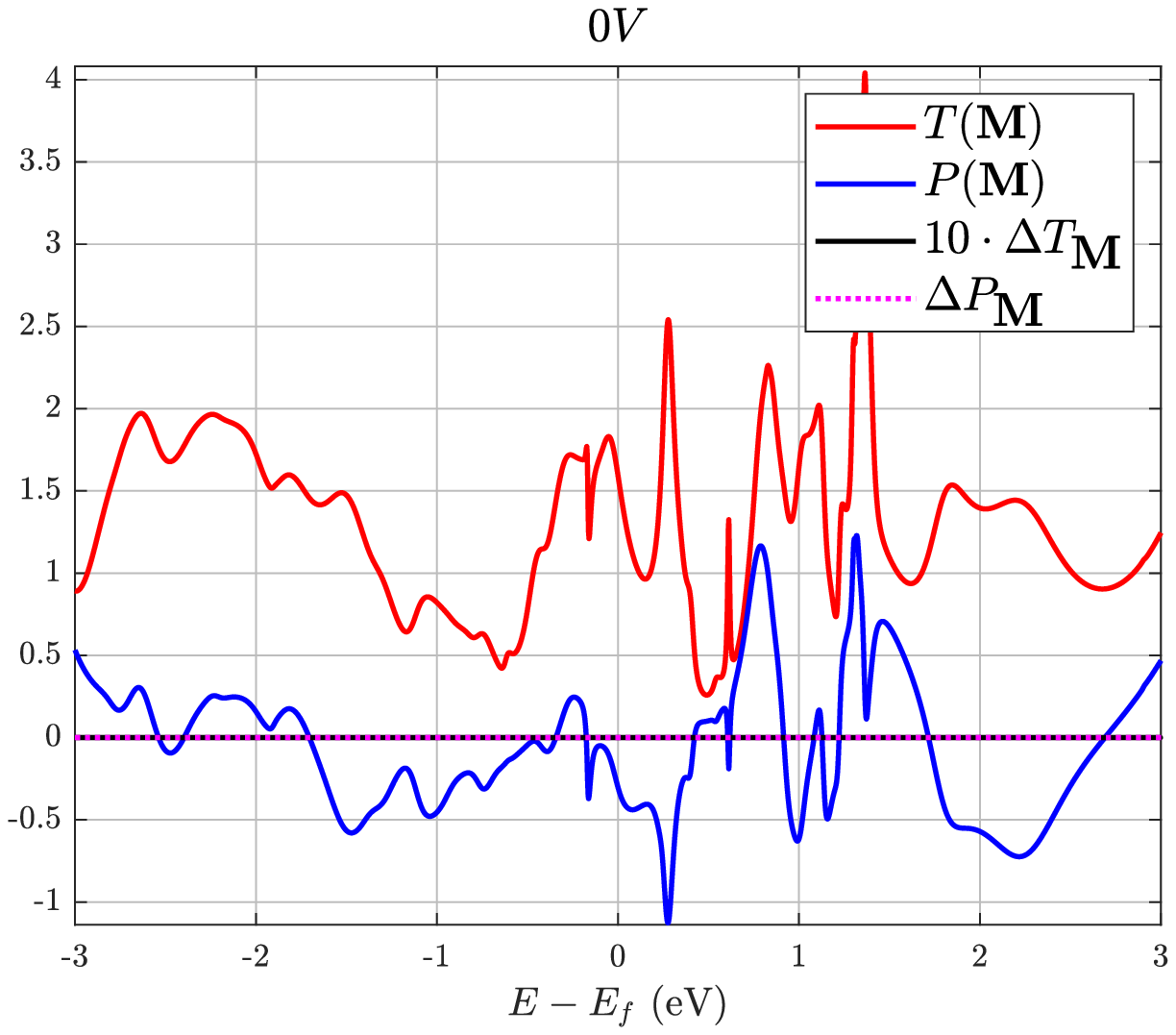}
\includegraphics[width=0.45\textwidth]{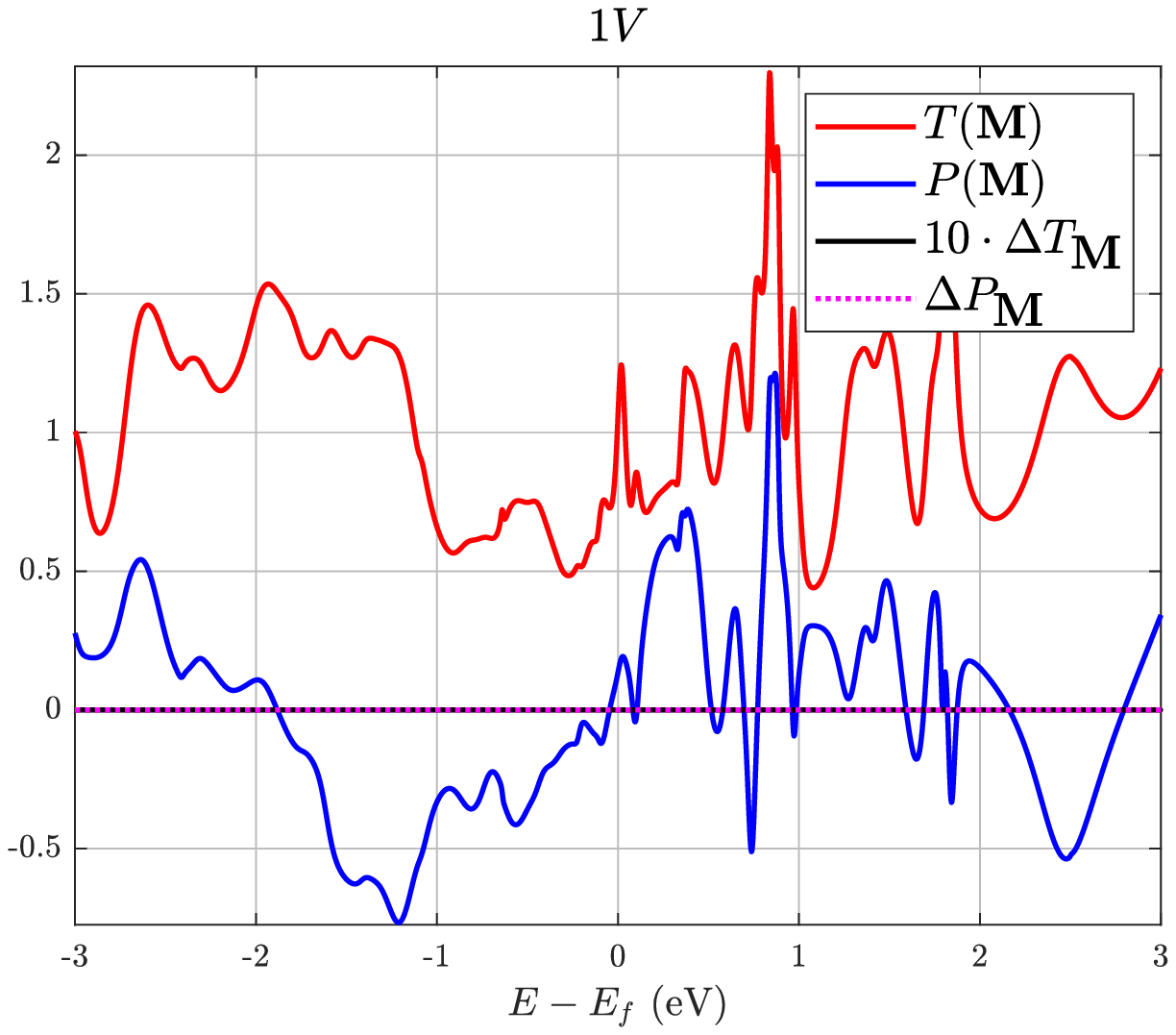}
\caption{
$\bm{M}=M\hat{\bm{x}}$ (longitudinal). Full symmetry: $\pazocal{G}=\left\{\begin{aligned}D_{4h}\text{, without mag.}&\\mm'm'\text{, with mag.}&\end{aligned}\right\}$}
\end{figure}

\begin{figure}[h!]
\centering 
\includegraphics[width=0.45\textwidth]{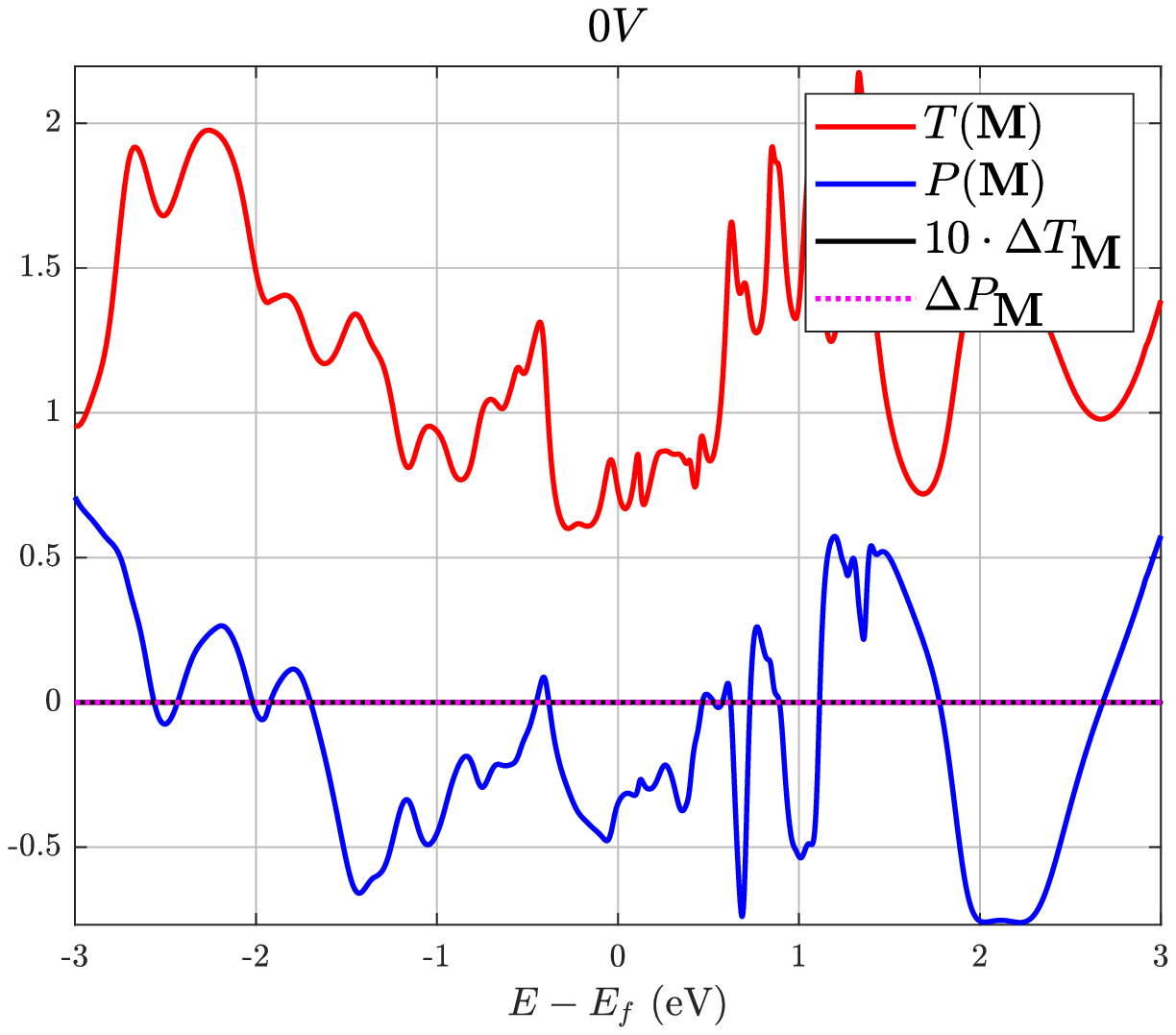}
\includegraphics[width=0.45\textwidth]{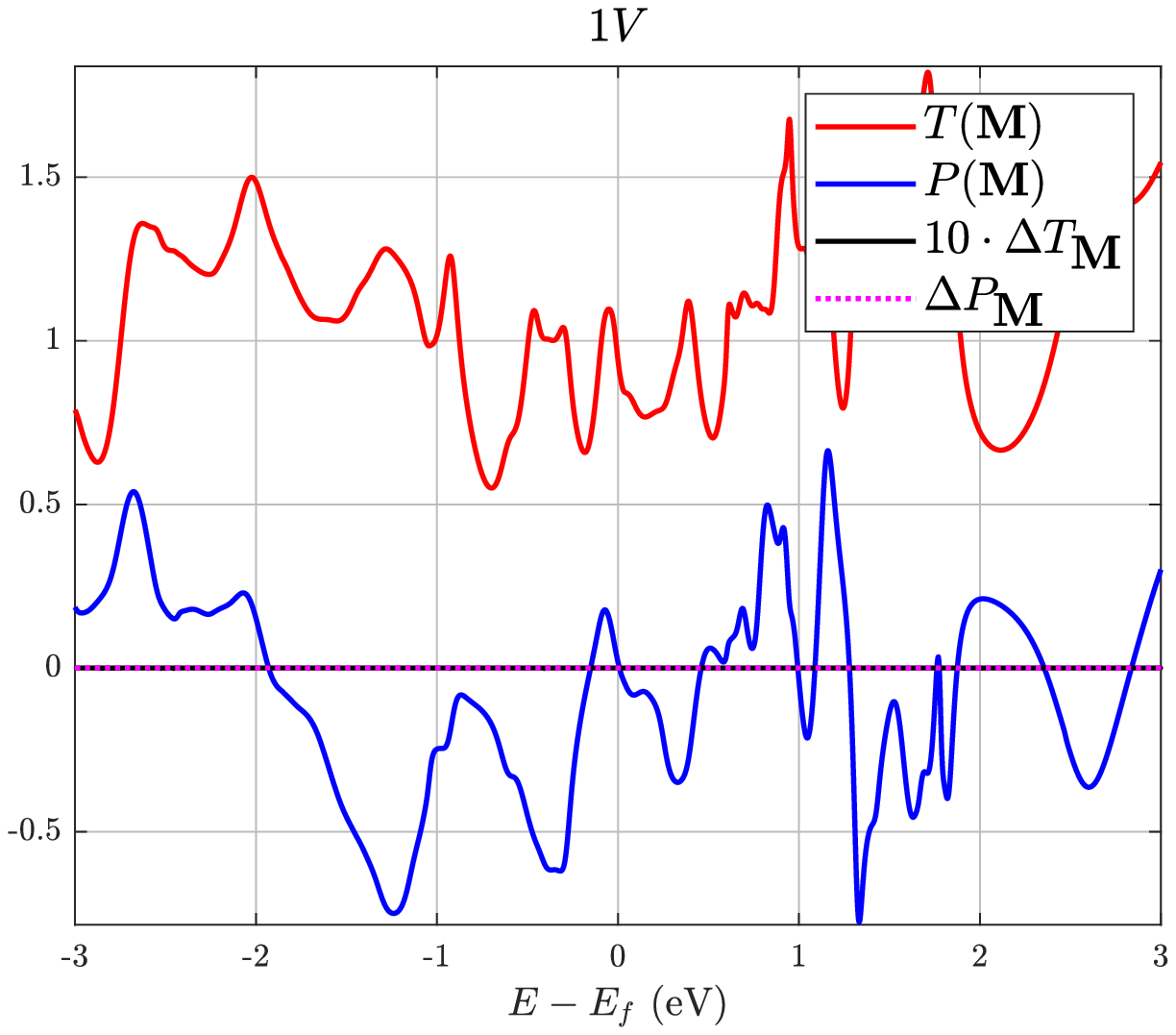}
\caption{
$\bm{M}=M\hat{\bm{z}}$ (transversal). Full symmetry: $\pazocal{G}=\left\{\begin{aligned}C_{1v}=\set{E,\sigma_{y}}\text{, without mag.}&\\m'=\set{E,\Theta\sigma_{y}}\text{, with mag.}&\end{aligned}\right\}$}
\end{figure}

\begin{figure}[h!]
\centering 
\includegraphics[width=0.45\textwidth]{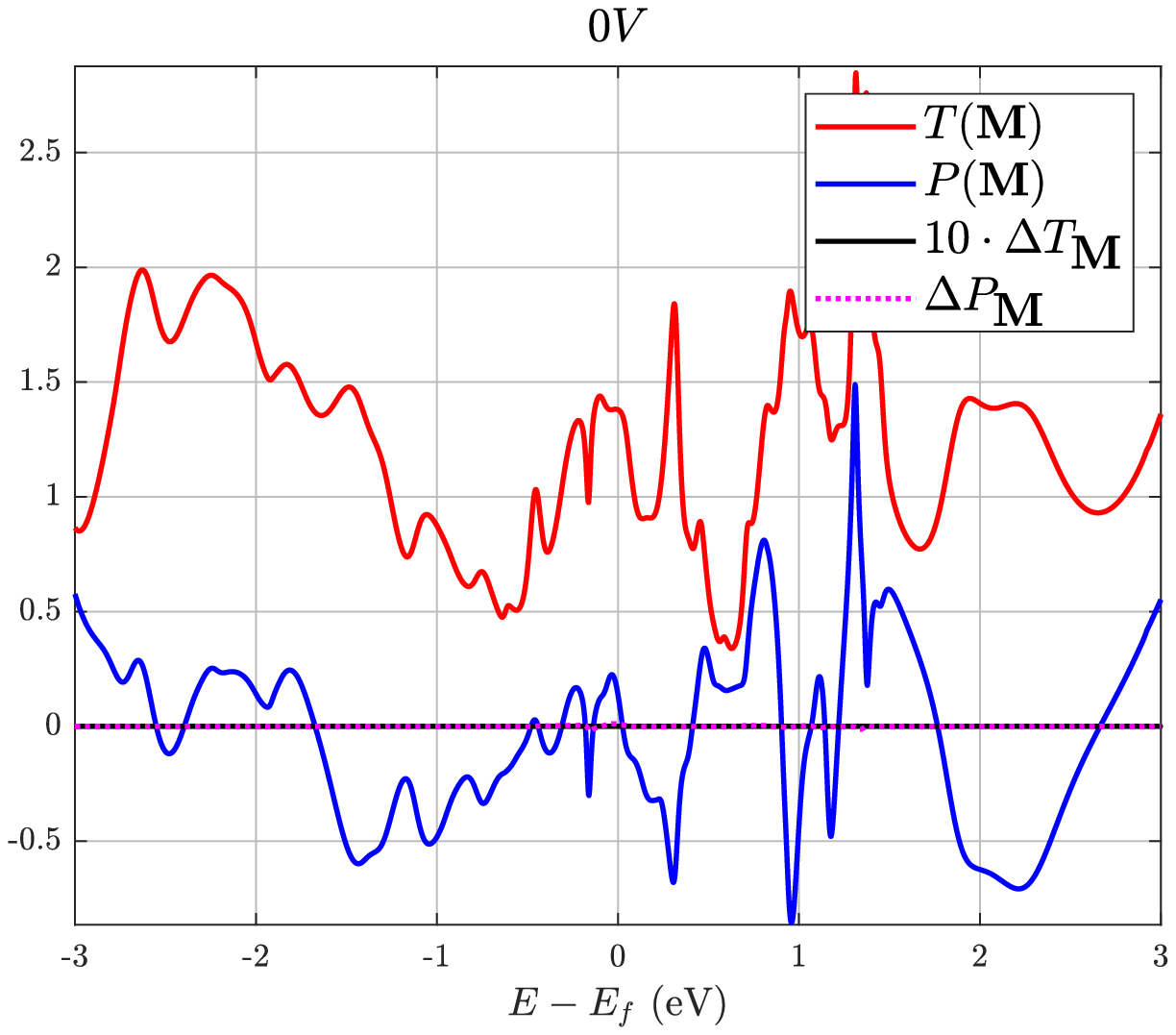}
\includegraphics[width=0.45\textwidth]{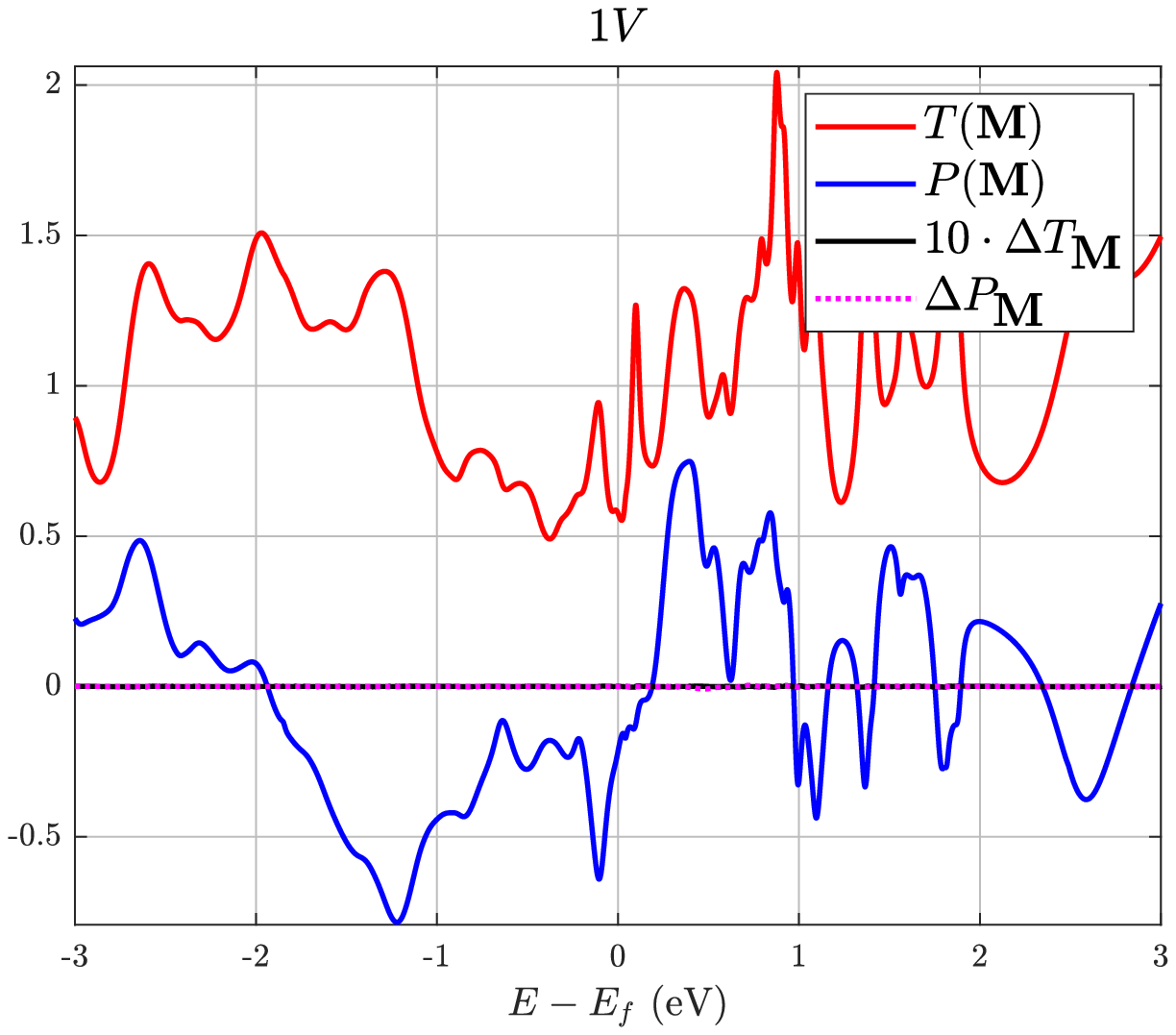}
\caption{
$\bm{M}=M\hat{\bm{x}}$ (longitudinal). Distorted: $\pazocal{G}=\left\{\begin{aligned}C_{1v}=\set{E,\sigma_{y}}\text{, without mag.}&\\m'=\set{E,\Theta\sigma_{y}}\text{, with mag.}&\end{aligned}\right\}$}
\end{figure}

\begin{figure}[h!]
\centering 
\includegraphics[width=0.45\textwidth]{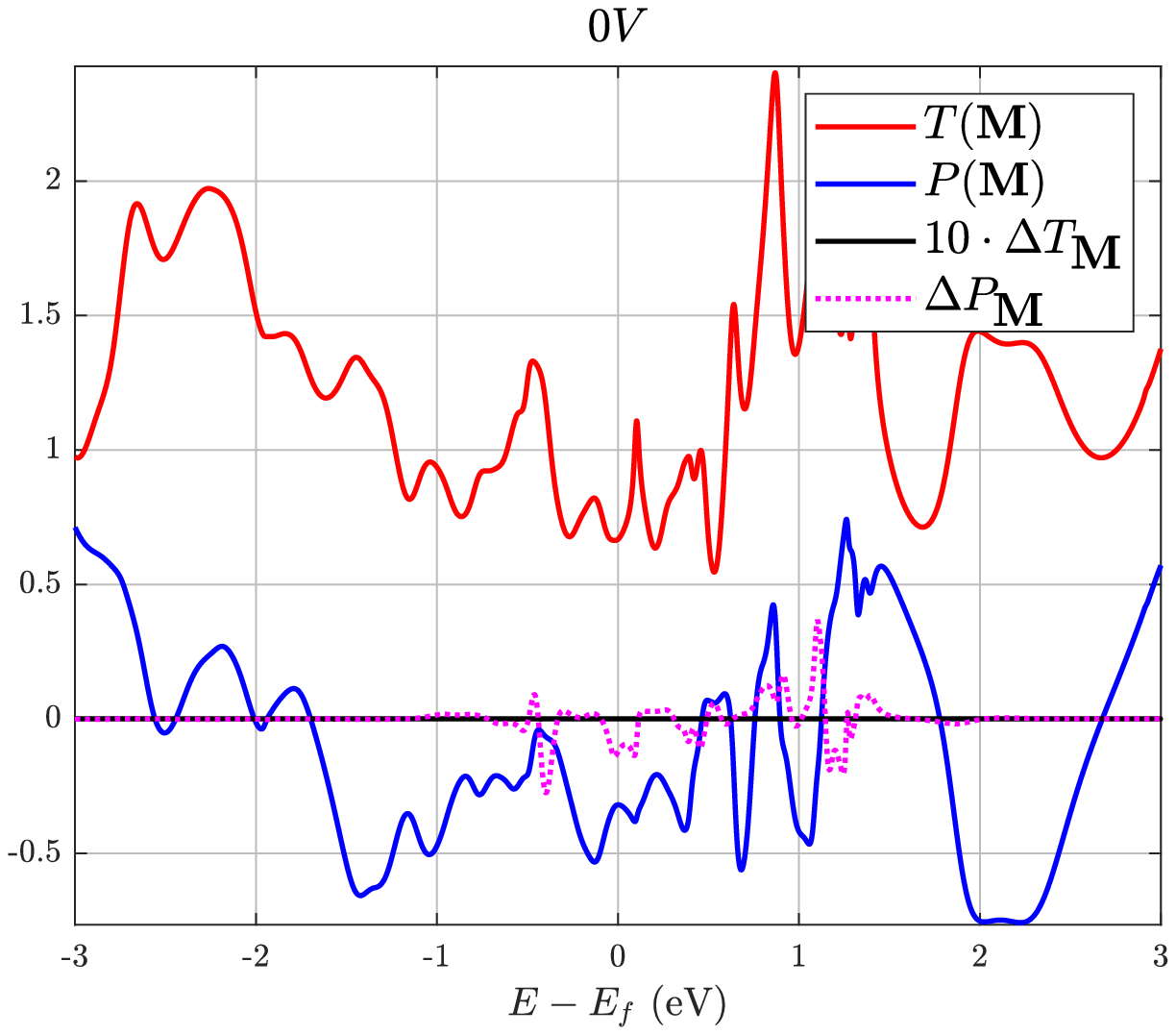}
\includegraphics[width=0.45\textwidth]{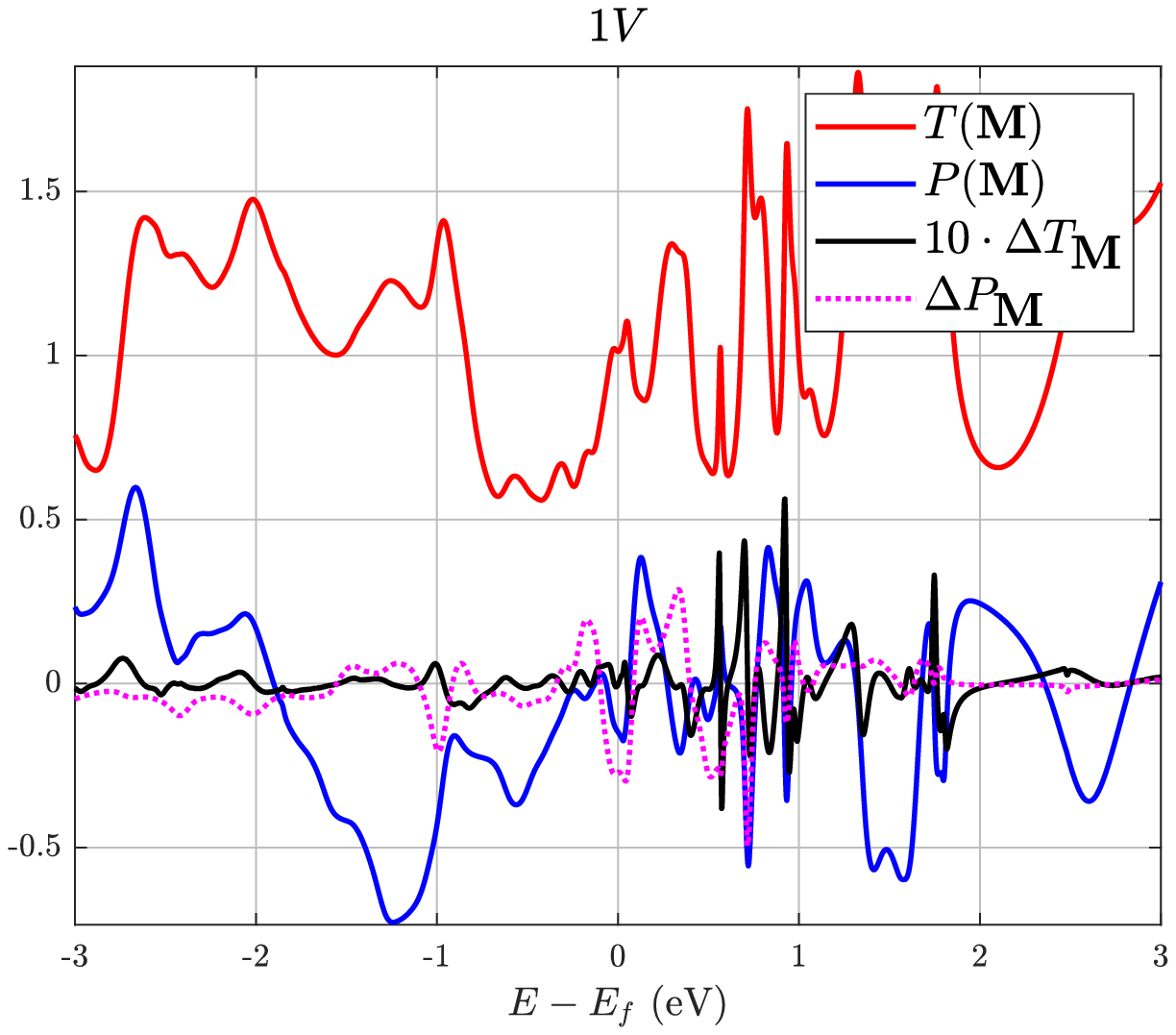}
\caption{
$\bm{M}=M\hat{\bm{y}}$ (transversal). Distorted: $\pazocal{G}=\set{E,\sigma_{y}}$}
\end{figure}

\begin{figure}[h!]
\centering 
\includegraphics[width=0.45\textwidth]{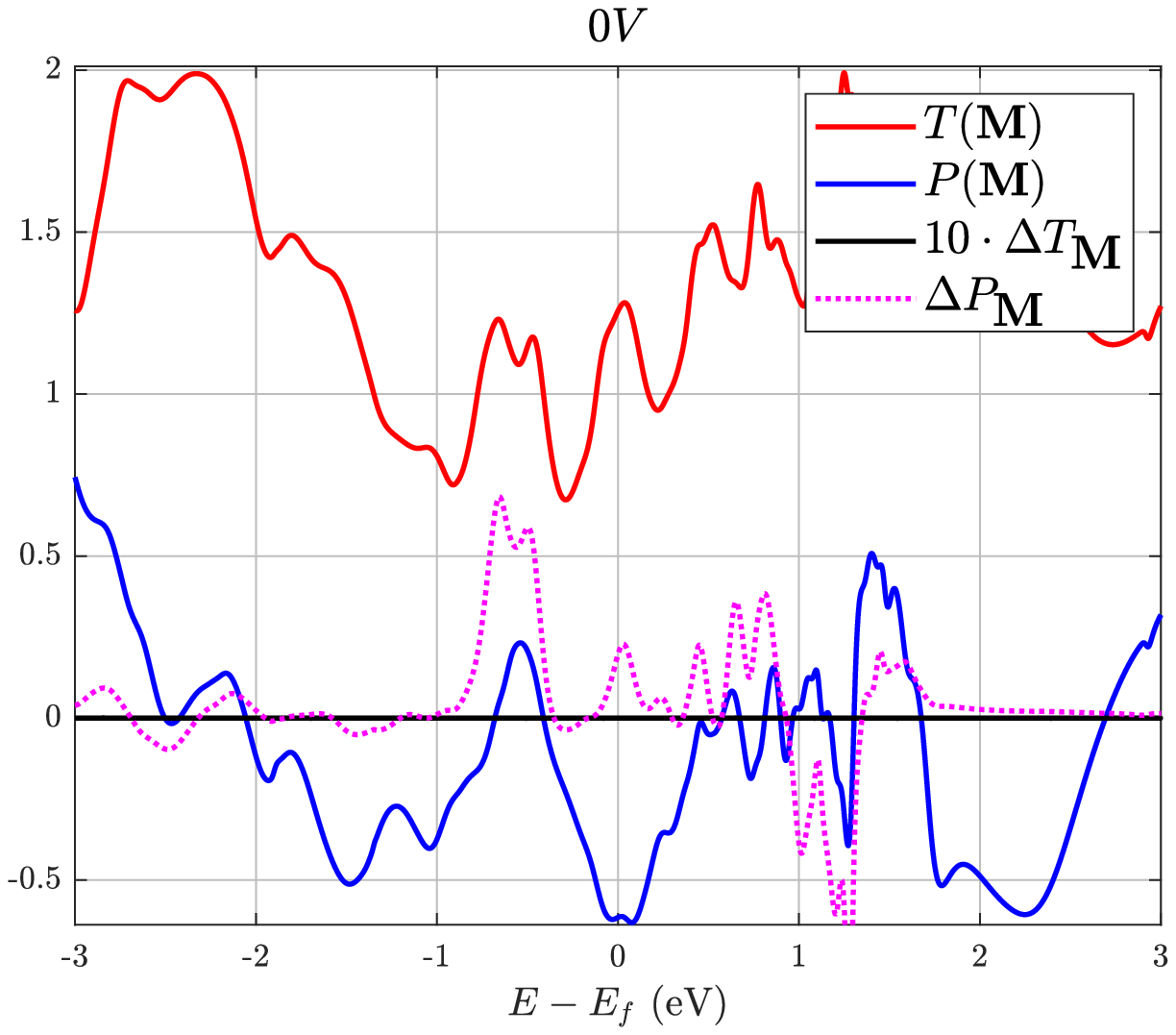}
\includegraphics[width=0.45\textwidth]{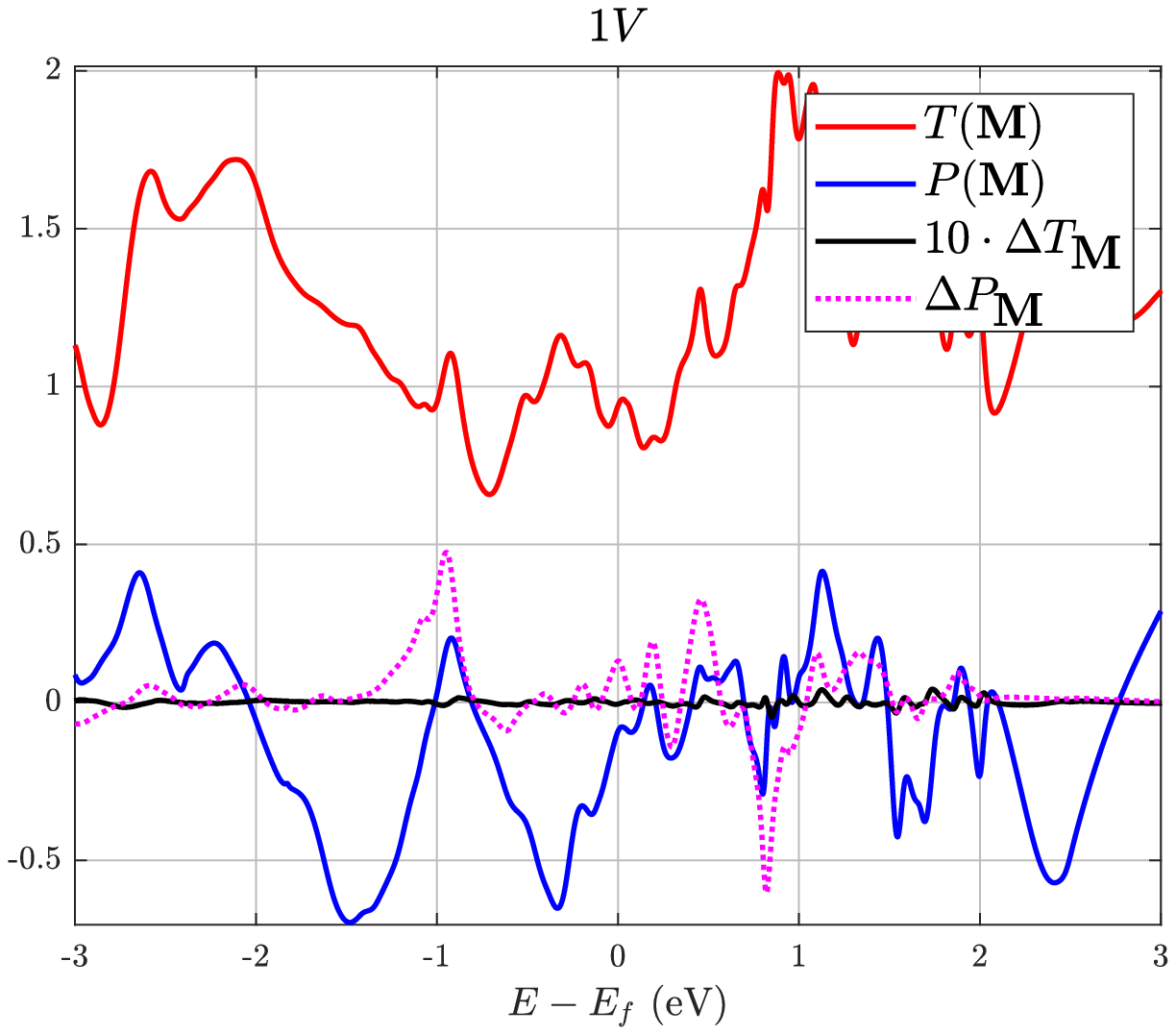}
\caption{
$\bm{M}=M\hat{\bm{z}}$ (transversal). Distorted: No symmetries}
\end{figure}

\begin{figure}[h!]
\centering 
\includegraphics[width=0.45\textwidth]{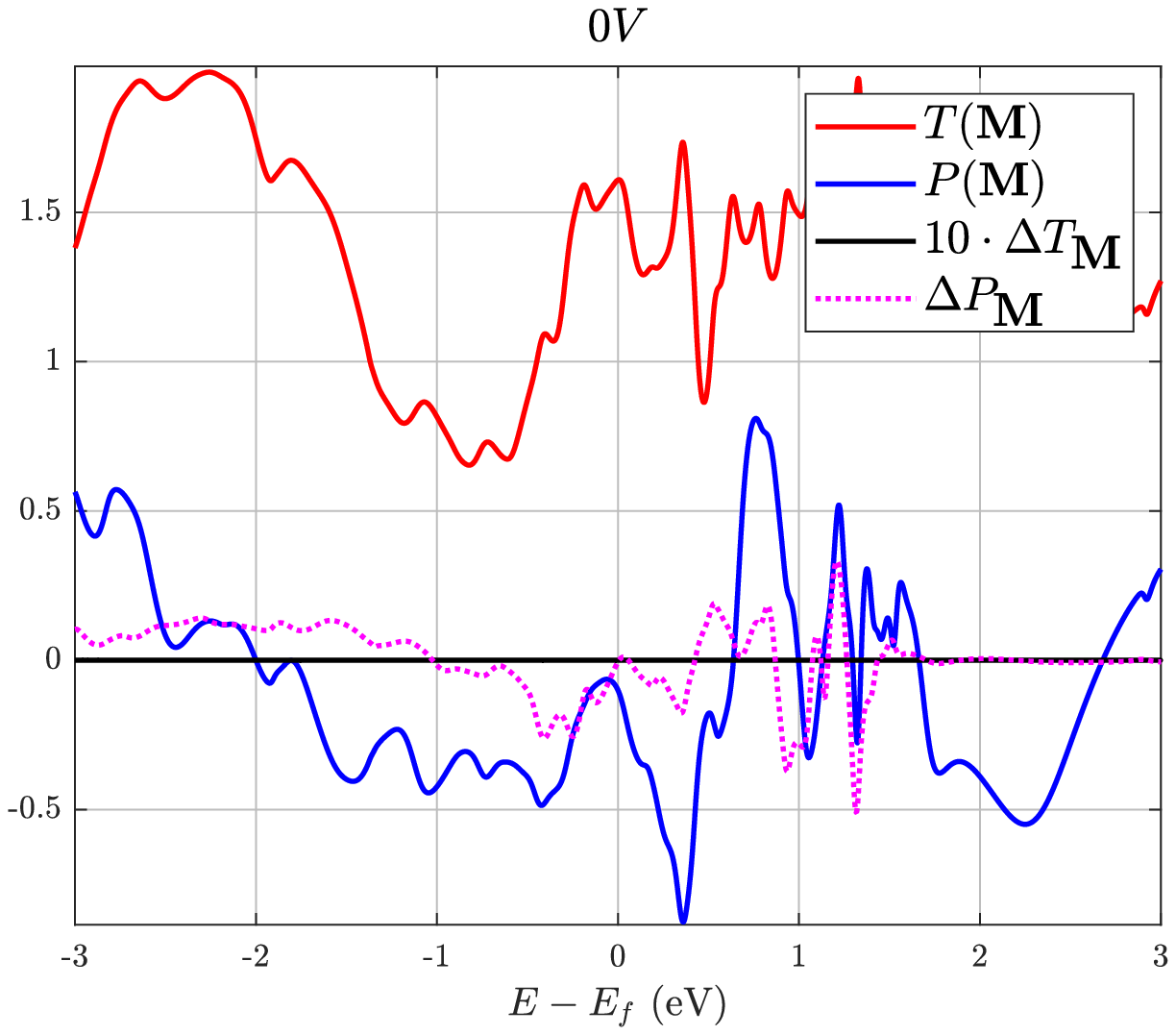}
\includegraphics[width=0.45\textwidth]{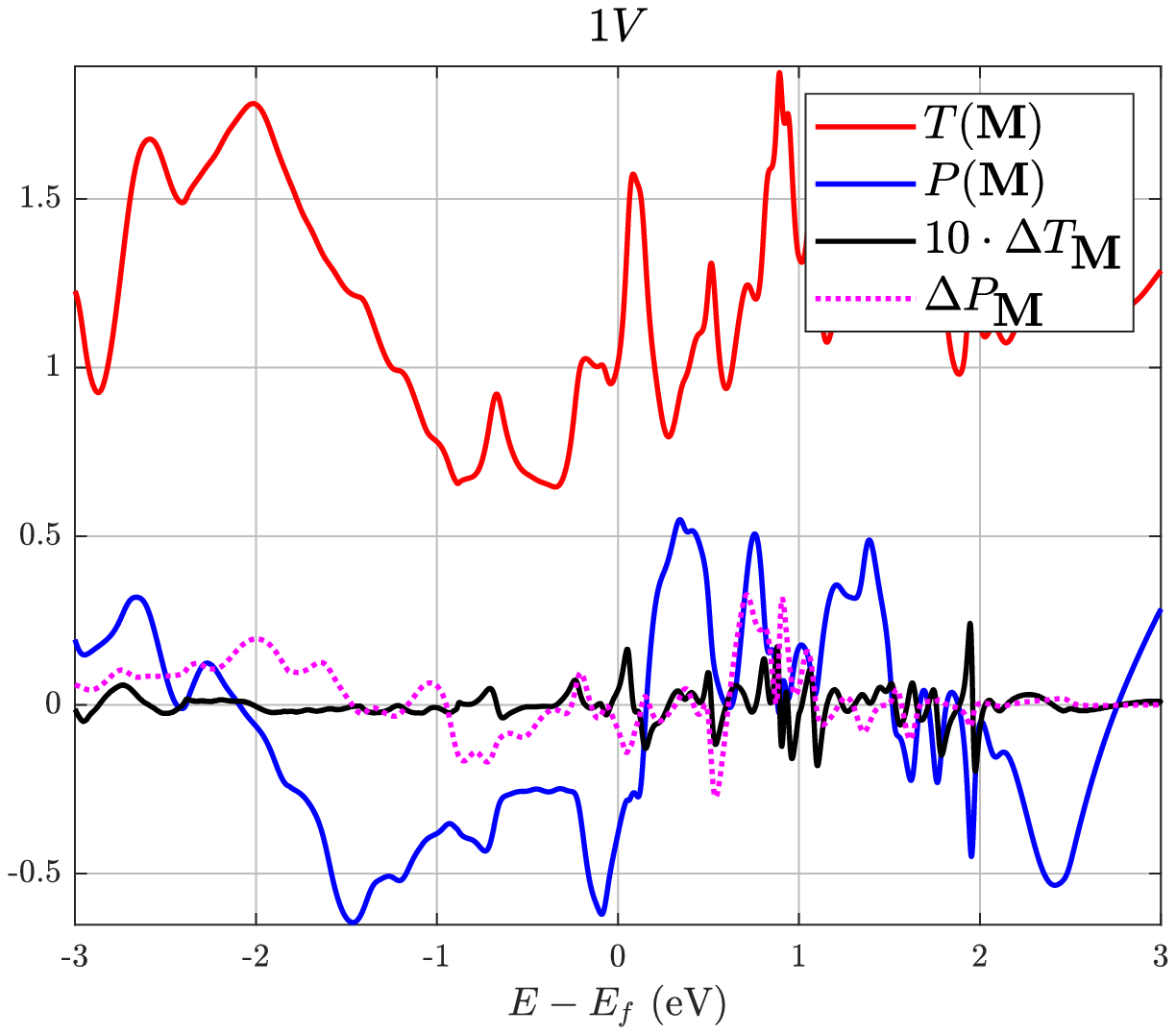}
\caption{
$\bm{M}=M\hat{\bm{x}}$ (longitudinal). Distorted: No symmetries}
\end{figure}

\begin{figure}[h!]
\centering 
\includegraphics[width=0.45\textwidth]{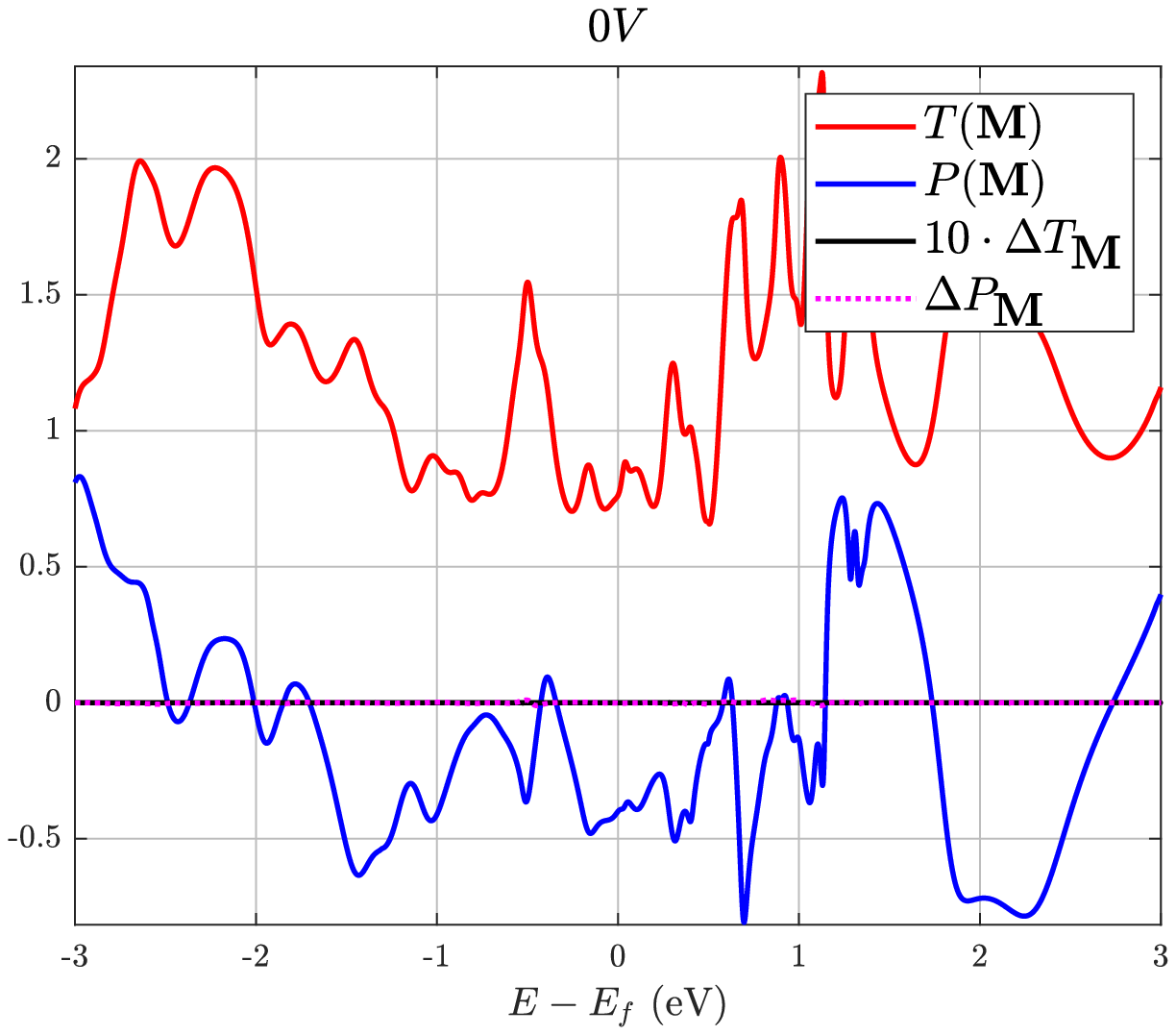}
\includegraphics[width=0.45\textwidth]{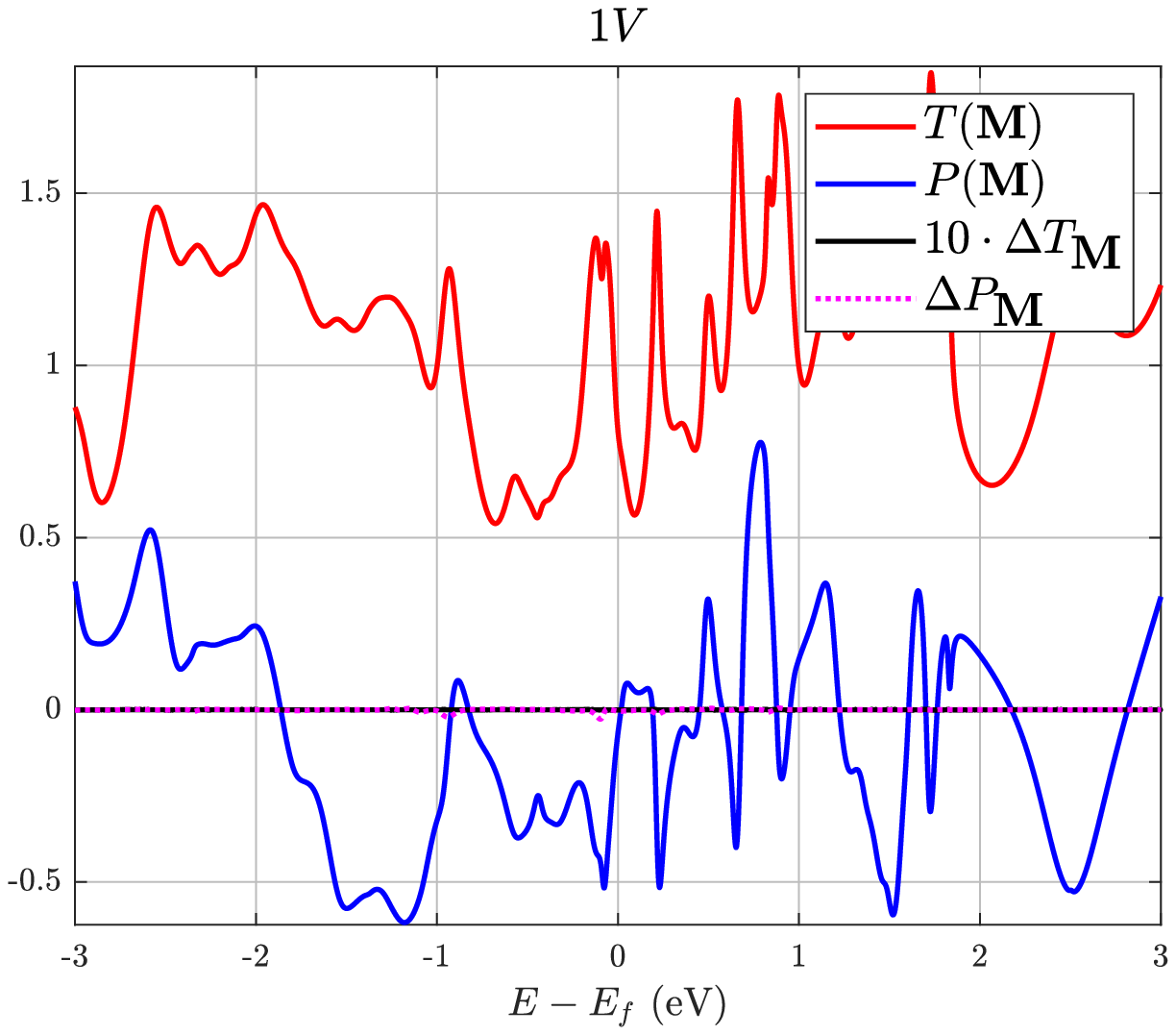}
\caption{
$\bm{M}=M\hat{\bm{z}}$ (transversal). Relative rotation of 15$^{\circ}$: $\pazocal{G}=\left\{\begin{aligned}C_{4}=\set{E,C_{4x},C_{2x},C^{-1}_{4x}}&\text{, without mag.}\\2'=\set{E,\Theta C_{2x}}&\text{, with mag.}\end{aligned}\right\}$}
\end{figure}

\clearpage
\subsection{W($C_{3}$)-W($C_{3}$)-Ni($C_{3}$) (30$^{\circ}$ relative rotation)}

\begin{figure}[h!]
\centering 
\includegraphics[width=0.6\textwidth]{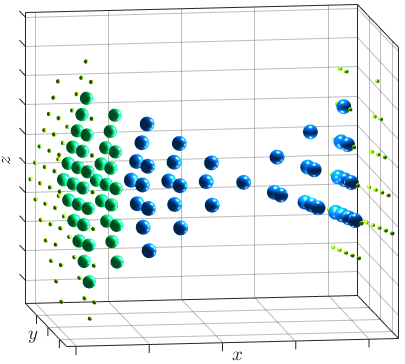}
\caption{}
\end{figure}

\begin{figure}[h!]
\centering 
\includegraphics[width=0.45\textwidth]{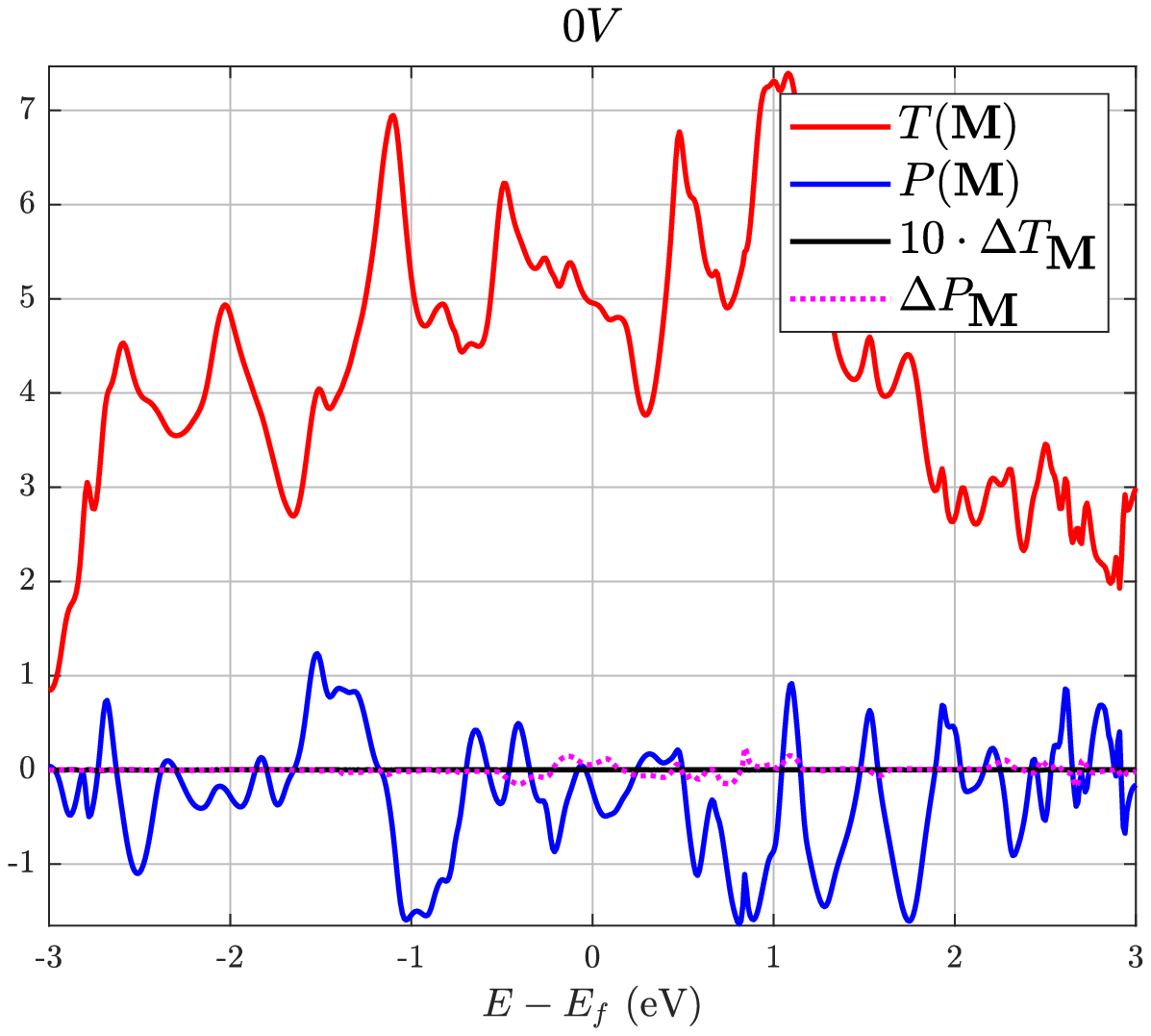}
\includegraphics[width=0.45\textwidth]{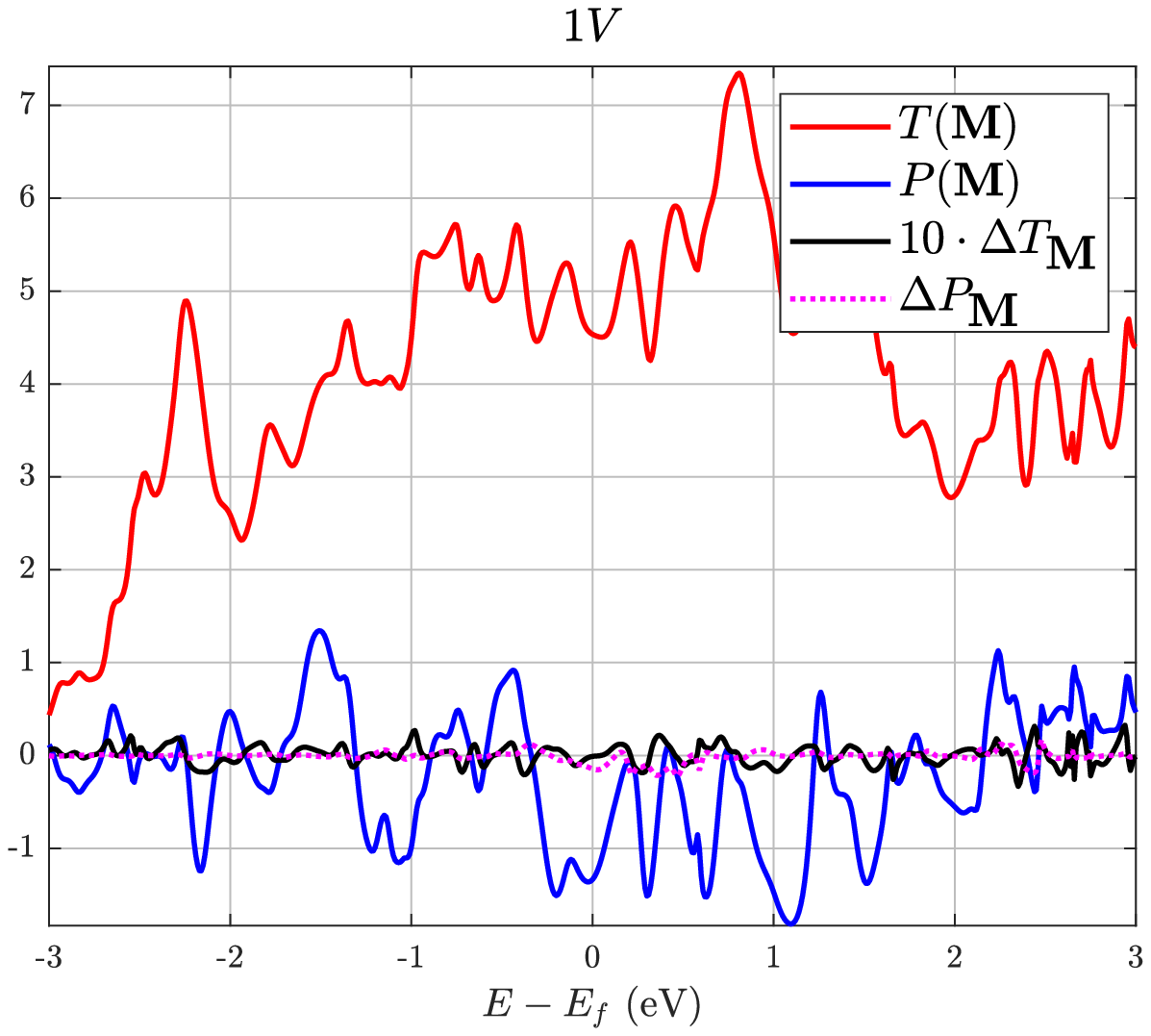}
\caption{
$\bm{M}=M\hat{\bm{z}}$ (transversal). $\pazocal{G}=\left\{\begin{aligned}&C_{3}=\set{E,C_{3x},C^{-1}_{3x}}\text{, without mag.}&\\&\text{Trivial, with mag.}&\end{aligned}\right\}$}
\end{figure}

\clearpage
\subsection{Pt($C_{4}$)-Helicene-Pt($C_{4}$)-Ni($C_{4}$)}

\begin{figure}[h!]
\centering 
\includegraphics[width=0.7\textwidth]{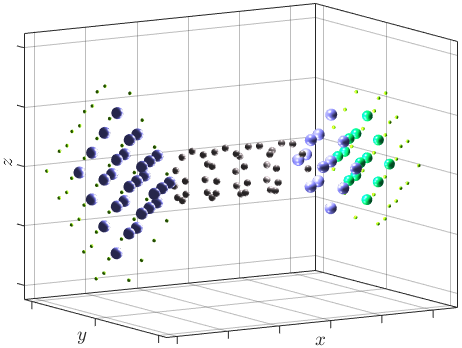}
\caption{}
\end{figure}

\begin{figure}[h!]
\centering 
\includegraphics[width=0.45\textwidth]{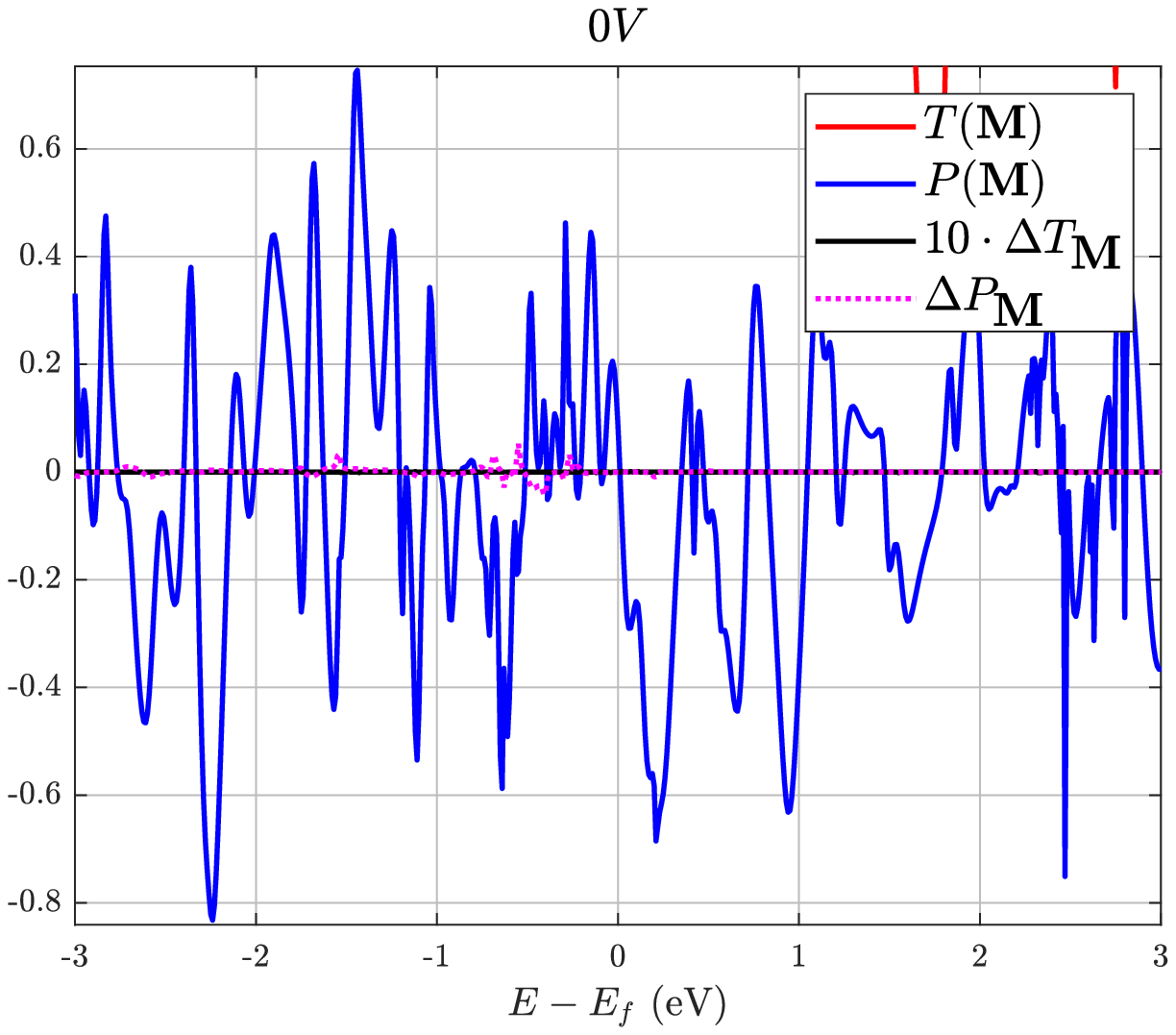}
\includegraphics[width=0.45\textwidth]{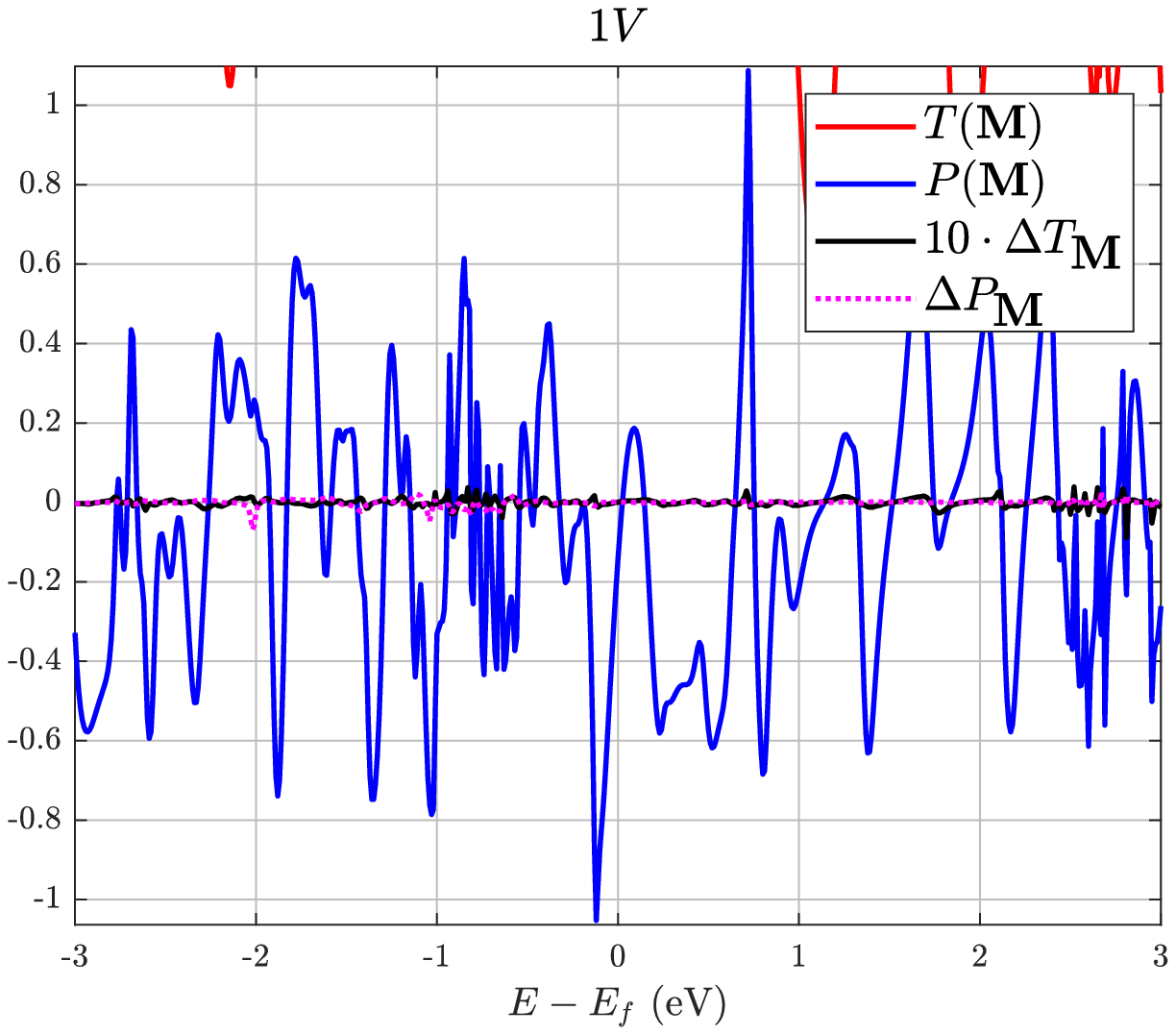}
\caption{
$\bm{M}=M\hat{\bm{z}}$ (transversal). No symmetries}
\end{figure}

